 \documentclass[10pt,aps,twocolumn,secnumarabic,amssymb, nobibnotes,  prx,superscriptaddress,floatfix]{revtex4-2}

\usepackage{graphicx}
\usepackage{dcolumn}
\usepackage{lipsum}

\usepackage{bm}
\usepackage[caption=false]{subfig}
\usepackage[colorlinks=true, citecolor=blue, linkcolor=blue, urlcolor=blue]{hyperref}
\usepackage{pinlabel} 
\usepackage{amsmath}
\allowdisplaybreaks
\usepackage{algorithm}
\usepackage{algpseudocode}
\usepackage{natbib}
\usepackage{braket}

\usepackage{amssymb}
\usepackage{mathrsfs}
\usepackage[T1]{fontenc}

\usepackage{color}

\begin{document}

\title{Parity Cross-Resonance: A Multi-qubit Gate}

\author{Xuexin Xu}
\affiliation{Peter Gr\"unberg Institute, Forschungszentrum J\"ulich, J\"ulich 52425, Germany}
\affiliation{J\"ulich-Aachen Research Alliance (JARA), Fundamentals of Future Information Technologies, J\"ulich 52425, Germany}

 \author{Siyu Wang}
\affiliation{Peter Gr\"unberg Institute, Forschungszentrum J\"ulich, J\"ulich 52425, Germany}
\affiliation{Institute for Quantum Information, RWTH Aachen University, D-52056 Aachen, Germany}
 
 \author{Radhika Joshi}
\affiliation{Peter Gr\"unberg Institute, Forschungszentrum J\"ulich, J\"ulich 52425, Germany}
\affiliation{Institute for Quantum Information, RWTH Aachen University, D-52056 Aachen, Germany}
 
 \author{Rihan Hai}
\affiliation{Department of Software Technology, Delft University of Technology, Delft, Netherlands}

 \author{Mohammad H. Ansari}
\affiliation{Peter Gr\"unberg Institute, Forschungszentrum J\"ulich, J\"ulich 52425, Germany}
\affiliation{J\"ulich-Aachen Research Alliance (JARA), Fundamentals of Future Information Technologies, J\"ulich 52425, Germany}

\begin{abstract}

The realization of multi-qubit entangling gates is essential for efficient, scalable, and fault-tolerant quantum information processing, reducing algorithmic complexity and circuit depth. We demonstrate a native three-qubit entangling gate implemented by simultaneously driving all qubits at a common frequency, exploiting engineered interactions to realize multi-control operations in a single coherent step. By optimizing the conditional dynamics originating from drive-induced nonlocal contamination, desired interaction channels are selectively enhanced while spurious terms are suppressed, ensuring robust performance within the computational subspace. This gate enables key applications, including deterministic GHZ-state generation, Toffoli-class logic with a shortest gate duration of 90 ns and a highest fidelity of 99.72\%, and a controlled-$ZZ$ gate tailored for fast surface-code quantum error correction. Simulations based on realistic IBM device parameters indicate that the gate maintains high fidelity and resilience under increasing excitation numbers and larger Hilbert-space dimensions. Our results establish a foundation for co-designing circuit architectures and control strategies that harness native multi-qubit interactions as fundamental building blocks for next-generation superconducting quantum processors, thereby enabling improved gate performance with more flexible tuning of circuit parameters.
\end{abstract}

\maketitle

\section{Introduction}

Multi-qubit entangling gates are essential for enhancing the efficiency and scalability of quantum information processing. Instead of decomposing multi-qubit operations into lengthy sequences of two-qubit gates~\cite{kandala21demonstration,sung21realization,kandala21demonstration,moskalenko22high}, higher-order entangling gates can directly generate complex correlations, such as Greenberger–Horne–Zeilinger (GHZ) states or the \emph{i}Toffoli gate, in a single step. These native multi-qubit interactions offer a compact representation of non-local transformations and enable more efficient implementations of quantum error-correcting codes, variational ans\"atze, and many-body simulations. Exploiting intrinsic multi-body couplings in platforms such as trapped ions~\cite{hahn19integrated,bruzewicz19dual-species,grzesiak20efficient,blumel21power-optimal}, Rydberg atoms~\cite{levine19parallel,khazali20fast,evered23high-fidelity,cao24multi-qubit}, or superconducting circuits~\cite{kim22high-fidelity,menke22demonstration,warren23extensive,xu24lattice} further improves fidelity and execution time, making the control of genuine multi-qubit entanglement a key milestone toward large-scale, fault-tolerant quantum computation.

Multipartite entangled states, such as GHZ, W, and Dicke states, serve as fundamental resources for a wide range of quantum information protocols, including quantum teleportation, superdense coding, and measurement-based quantum computation. The capability to efficiently generate, manipulate, and preserve such entanglement is therefore essential for achieving scalable and fault-tolerant quantum technologies. Recent experimental progress includes the generation of a 10-qubit W state with a fidelity of 91.3\% and a double-excitation Dicke state with a fidelity of 92.6\%~\cite{chen25efficient},  as well as the reported largest 120-qubit GHZ state with a fidelity of 56\% demonstrated on IBM quantum processors~\cite{javadi-abhari25big-cats}.

Direct multi-qubit gates can substantially reduce circuit depth, thereby mitigating cumulative errors, minimizing decoherence, and enhancing computational efficiency. One approach is to synchronize interactions, for example, aligning spin-orbit coupling and orbital magnetic fields to realize a $C^2R_y(\pi)$ gate on spin-qubit platforms~\cite{nguyen25single-step}. Another practical route is through simultaneous two-qubit operations, such as the controlled-controlled-$Z$ and controlled-\emph{i}SWAP gates, realized via parallel two-qubit operations on different qubit pairs~\cite{gu21fast,tasler25optimizing,old25fault-tolerant} on superconducting circuits, as well as the one-step three-qubit parity gate~\cite{itoko24three-qubit} and the \emph{i}Toffoli gate~\cite{kim22high-fidelity}, implemented via concurrent cross-resonance (CR) microwave drives applied to all qubits~\cite{rigetti10fully,chow11simple,chow12universal,corcoles13process,sheldon16procedure,magesan20effective}. Typically, these multi-qubit gates would require at least six two-qubit CNOT-like operations if decomposed sequentially; leveraging the simultaneous-gate approach markedly reduces gate count and execution time, thereby improving overall circuit robustness and efficiency.

However, simultaneous-gate approaches trade off unwanted parasitic couplings to enable the joint implementation of multiple gates, limiting achievable fidelities and demanding careful circuit optimization. To overcome these challenges, we develop and benchmark a general protocol for synthesizing a single-shot parity cross-resonance (PCR) gate characterized by a direct three-qubit $ZZX$ interaction in fixed-frequency transmon circuits~\cite{ansari0three}. In contrast to Refs.~\cite{itoko24three-qubit,kim22high-fidelity}, which rely on simultaneous two-qubit interactions, this method combines perturbative design with gradient-free, non-perturbative optimization to realize a one-step three-body interaction beyond the dispersive regime, while automatically identifying optimal static operating parameters and suppressing stray interactions. Applied to a representative three-qubit layout based on IBM parameters, the protocol identifies broad operating regions supporting strong $ZX$-like rates, enabling high-fidelity GHZ-state preparation and single-shot CCNOT and \emph{i}Toffoli gate implementations. It also extends naturally to quantum error correction by adapting the controlled-$ZZ$ (CZZ) gate for direct parity mapping, enhancing both speed and fidelity.

The paper is organized as follows: In Sec.~\ref{sec:principle}, we outline the principles of the three-body interaction quantum logic gate and its applications to GHZ state preparation (Sec.~\ref{subsec:principle_ghz}), CCNOT-like implementation (Sec.~\ref{subsec:principle_ccnot}), and error-correction–oriented controlled-$ZZ$ gates  (Sec.~\ref{subsec:principle_czz}), followed by their individual implementations  (Sec.~\ref{subsec:principle_implementation}).
In Sec.~\ref{sec:optimization}, we describe our gradient-free optimization framework for modeling and tuning the gate parameters.  
Finally, Sec.~\ref{sec:validation} presents performance validation for the three main examples: GHZ triplet state (Sec.~\ref{subsec:ghz_validation}), CCNOT-like gate (Sec.~\ref{subsec:ccnot_validation}), and CZZ gate (Sec.~\ref{subsec:czz_validation}).

\section{Principle}\label{sec:principle}
The $ZZX$ interaction, typically arising as a higher-order by-product of two simultaneous CR gates, is often neglected in standard gate synthesis procedures. Yet, under specific operating conditions---detailed later in the manuscript---this three-body term can be significantly amplified. When acting on three qubits with all-to-all connectivity, the $ZZX$ interaction implements a $Z$-parity check between the first two qubits while simultaneously flipping the state of the third. This flexibility makes the $ZZX$ interaction a valuable fundamental building block for quantum error detection schemes~\cite{tasler25optimizing}, and leads us to introduce it as the \textit{Parity Cross-Resonance} (PCR) gate.

Native multi-qubit interactions constitute a foundational asset in quantum logic synthesis, as they permit the direct implementation of otherwise composite operations---operations that, in their conventional form, require sequences of two-qubit gates interspersed with single-qubit rotations. Harnessing such intrinsic couplings not only compresses circuit depth but also suppresses spurious Hamiltonian terms, thereby improving both the fidelity and resilience of entanglement generation on NISQ-era superconducting platforms. Inspired by these benefits, we now proceed to investigate how the PCR gate can be leveraged to streamline quantum circuit design and to realize efficient, hardware-native multi-qubit operations.

\subsection{Quantum Logic}
We now present practical protocols that exploit the $U_{ZZX}(\theta)$ interaction to enable efficient entanglement generation via the PCR gate. The associated unitary action on computational basis states $\ket{q_1 q_2 q_3}(q_1, q_2, q_3\in\{0,1\})$ is given by
\begin{eqnarray}
\!\!\!\!\! U_{ZZX}({\theta}) \ket{q_1 q_2 q_3} && = \cos\left(\frac{\theta}{2}\right)\ket{q_1 q_2 q_3}  \nonumber \\ &&   - i(-1)^{q_1\oplus q_2} \sin\left(\frac{\theta}{2}\right)\ket{q_1 q_2 \overline{q_3}},
\end{eqnarray}
For the case of $\pi$ rotation, the operator returns the following values: 
\[
U_{ZZX}\left({\pi}\right) \ket{q_1 q_2 {q}_3}  = 
\begin{cases}
- i \ket{q_1 q_2 \overline{q_3}} & \text{if } q_1 = q_2 \\
+ i \ket{q_1 q_2 \overline{q_3}} & \text{if } q_1 = \overline{q_2}
\end{cases}
\]

A $2\pi$ rotation returns the state back on itself, which indicates that by negating the rotation angle $\pi \to -\pi $ the gate is inverted. Moreover, the gate at $\pi/2$ rotation creates a conditional superposition on the third qubit:
\[
U_{ZZX}\left(\frac{\pi}{2}\right) \ket{q_1 q_2 {q}_3}  = 
\begin{cases}
\frac{1}{\sqrt{2}}\ket{q_1 q_2} \left(\ket{q_3} - i \ket{\overline{q_3}}\right) & \text{if } q_1 = q_2 \\
\frac{1}{\sqrt{2}}\ket{q_1 q_2} \left(\ket{q_3} + i \ket{\overline{q_3}}\right)& \text{if } q_1 = \overline{q_2}
\end{cases}
\]

These capabilities are particularly valuable for three key purposes: (1) the preparation of Greenberger–Horne–Zeilinger (GHZ) entangled states, and (2) the implementation of a CCNOT-like multi-qubit gate, (3) the realization of controlled-$ZZ$ (CZZ) gate.

\subsubsection{GHZ State Preparation}\label{subsec:principle_ghz}

A GHZ state is a maximally entangled multi-qubit state of the form  $|\text{GHZ}\rangle = \frac{1}{\sqrt{2}} \left( |000\cdots\rangle + |111\cdots\rangle \right)$. Such  states serve as fundamental resources in quantum information processing~\cite{hillery99quantum}, playing critical roles in non-locality tests~\cite{buhrman10nonlocality}, quantum error correction~\cite{divincenzo96fault-tolerant,preskill98reliable}, and entanglement-based protocols~\cite{bose98multiparticle,yang04efficient}. Mitigating errors during GHZ state preparation involves reducing the effects of decoherence, gate imperfections, and crosstalk encountered during entangling operations. Recent advancements include the use of quantum autoencoders, which are trained to compress and reconstruct entangled input states, offering a data-driven approach to error mitigation~\cite{pazem25error}.

Since the $ZZX$ interaction can be regarded as a three-qubit analog of the two-qubit $ZX$ gate, we begin by recalling the preparation of a Bell state using the CR mechanism. Bell states constitute a subset of maximally entangled two-qubit states and admit natural generalizations to multipartite settings, such as GHZ states. The two-qubit $Z_1X_2$ interaction of CR gate enables the creation of the maximally entangled Bell state via the following gate sequence:
$ S_2 H_1 U_{Z_1 X_2}({\pi/2}) H_1 \ket{00} = \left({\ket{00} + \ket{11}}\right)/{\sqrt{2}}$,
where $S_i$  and $H_i$ represent the phase and  Hadamard gate, respectively, acting on qubit $i$, and the $U_{ZX}$ rotation is defined as $U_{Z_1X_2}(\theta) = \cos\left({\theta}/{2}\right) I_1I_2 - i \sin\left({\theta}/{2}\right) Z_1X_2$.

By analogy with the two-qubit CR gate, the three-qubit $ZZX$ interaction can be harnessed to directly generate a GHZ state. In particular, we consider the following protocol.
$S_3 H_2 H_1 U_{Z_1 Z_2 X_3}({{\pi}/{2}}) H_1 H_2 \ket{000} = \left({\ket{000} + \ket{111}}\right)/{\sqrt{2}}$.
Here, the central operation---$U_{Z_1 Z_2 X_3}({{\pi}/{2}})$---corresponds to evolution under the $ZZX$ Hamiltonian for a time corresponding to a $\pi/2$ rotation. The gate sequence $H_i Z_i H_i$ transforms the $Z_i$ operator into $X_i$, and the inclusion of the phase gate $S_i$ compensates for the acquired phase. This composite circuit consists of five single-qubit gates (typically ultrafast and high-fidelity) in conjunction with a single three-qubit entangling gate.  The resulting operation effectively realizes an interaction $U_{XXX}({{\pi}/{2}})$, which entangles all three qubits and maps the initial product state $\ket{000}$ to a GHZ state. This construction highlights the utility of native three-body interactions in reducing circuit complexity. Compared to the conventional GHZ state preparation protocol---typically requiring two CNOT gates---the $ZZX$-based protocol provides a compact alternative by replacing both entangling gates with a \emph{single} three-qubit unitary. The prerequisite for this replacement is that the three-qubit gate is applied in one go on three all-in-all interacting qubits, thus the chain structure is not favorable.

\subsubsection{A Three-Qubit CCNOT}\label{subsec:principle_ccnot}

Another natural application of the PCR gate is the implementation of multi-qubit logic gates. In particular, the three-qubit Toffoli gate---also known as the CCNOT gate---plays a central role in quantum information processing. As a universal gate for reversible classical computation, it enables the realization of arbitrary Boolean functions on quantum registers~\cite{nielsen02quantum}. Conventional synthesis of Toffoli-like gates typically requires at least $2n$ CNOT gates, resulting in substantial resource overhead~\cite{shende08on-the-cnot-cost}. Recent developments have shown that simultaneous application of two CR gates combined with the Pancharatnam-Berry geometric phase~\cite{cho19emergence}, can produce more efficient gate constructions with durations as short as 353~ns~\cite{kim22high-fidelity}. As an alternative, directly harnessing the native PCR interaction based on $ZZX_{\pi/4}$, supplemented by two-qubit $ZX$ operations, offers a compelling route toward reducing CNOT gate depth. Within this framework, the ideal Hamiltonian for realizing a CCNOT-like gate can be expressed as
\begin{align}\label{eq:ccnot}
H_{\text{CCNOT}} &= -\frac{\pi}{8}\,(I_1-Z_1)(I_2-Z_2)(I_3-X_3) \nonumber\\
&= \frac{\pi}{8}\,\big[(IZI + ZII + IIX) + ZZX \nonumber\\
&\quad - (ZZI + IZX + ZIX + III)\big] \nonumber\\
&\simeq \begin{pmatrix}
1 & 0 & 0 & 0 & 0 & 0 & 0 & 0 \\
0 & 1 & 0 & 0 & 0 & 0 & 0 & 0 \\
0 & 0 & 1 & 0 & 0 & 0 & 0 & 0 \\
0 & 0 & 0 & 1 & 0 & 0 & 0 & 0 \\
0 & 0 & 0 & 0 & 1 & 0 & 0 & 0 \\
0 & 0 & 0 & 0 & 0 & 1 & 0 & 0 \\
0 & 0 & 0 & 0 & 0 & 0 & 0 & 1 \\
0 & 0 & 0 & 0 & 0 & 0 & 1 & 0
\end{pmatrix}.
\end{align}

In this construction, all contributing Pauli terms commute, enabling coherent evolution under their simultaneous action. Applying the associated unitary operator to a computational basis state $\ket{q_1q_2q_3}$ yields:
\[
U(H_{\text{CCNOT}}) \ket{q_1q_2q_3} = e^{i m \frac{\pi}{8}} e^{-i m \frac{\pi}{8} X_3} \ket{q_1q_2q_3},
\]
where the phase coefficient is given by $m = 1 + (-1)^{q_1+q_2} - (-1)^{q_1}- (-1)^{q_2}$. It is straightforward to verify that $m = 4$ if and only if $q_1 = q_2 = 1$, corresponding to control state $\ket{11}$; for all other control states ($\ket{00}$, $\ket{01}$, $\ket{10}$), one finds $m = 0$. Hence, $m$ can be equivalently written as $4q_1q_2$, leading to the simplified unitary transformation:
\[
U_{\text{CCNOT}} \ket{q_1q_2q_3} = (1 - q_1q_2) \ket{q_1q_2q_3} +\, q_1q_2 \ket{q_1q_2\overline{q_3}}.
\]
This expression shows that for all control states except $\ket{11}$, the unitary reduces to identity. When $q_1 = q_2 = 1$, the third qubit flips, thereby implementing the standard CCNOT operation. To realize this gate via Hamiltonian engineering, one can construct an effective three-qubit interaction satisfying:
\[
\alpha_{ZZI} = \alpha_{IZX} = \alpha_{ZIX} = -\alpha_{ZZX}.
\]
Under this condition, the CCNOT gate is implemented as a single-shot evolution under the resulting effective Hamiltonian, possibly supplemented by local single-qubit rotations.

In many practical scenarios, it is advantageous to consider a variant known as the \emph{i}Toffoli gate, which omits stray $ZZ$ interactions and admits a simpler form:
\begin{align}
H_{i\text{Toffoli}} =& \frac{\pi}{8} \left[(IIX + ZZX) - (IZX + ZIX)\right]\nonumber\\
\simeq& \begin{pmatrix}
1 & 0 & 0 & 0 & 0 & 0 & 0 & 0 \\
0 & 1 & 0 & 0 & 0 & 0 & 0 & 0 \\
0 & 0 & 1 & 0 & 0 & 0 & 0 & 0 \\
0 & 0 & 0 & 1 & 0 & 0 & 0 & 0 \\
0 & 0 & 0 & 0 & 1 & 0 & 0 & 0 \\
0 & 0 & 0 & 0 & 0 & 1 & 0 & 0 \\
0 & 0 & 0 & 0 & 0 & 0 & 0 & i \\
0 & 0 & 0 & 0 & 0 & 0 & i & 0
\end{pmatrix}.
\end{align}
Applying the associated unitary yields:
\[
U_{i\text{Toffoli}} \ket{q_1q_2q_3} = (1 - q_1q_2) \ket{q_1q_2q_3} - i\, q_1q_2 \ket{q_1q_2\overline{q_3}}.
\]
As before, this expression implies that the gate acts as identity unless both control qubits are in the $\ket{1}$ state. In the $\ket{11}$ configuration, the third qubit undergoes a bit flip with an additional $-i$ phase, consistent with the definition of an \emph{i}Toffoli operation.

\subsubsection{A CZZ Gate for Surface-Code QEC}\label{subsec:principle_czz}
Quantum error correction (QEC) is essential for achieving fault-tolerant quantum computation.  
In surface-code architectures, stabilizer measurements are typically implemented through sequences of two-qubit gates, which can limit operation speed and introduce additional errors.  
Recent advances have proposed a three-qubit controlled-$Z$-$Z$ (CZZ) gate that enables direct parity mapping of two data qubits onto a measurement qubit in a single step, thereby achieving higher error thresholds and lower logical error rates compared to the standard CZ-based readout~\cite{tasler25optimizing,old25fault-tolerant}.  
In addition to its interpretation as two simultaneous CZ gates, the CZZ operation can also be realized via a direct three-body interaction, with the ideal Hamiltonian given by
\begin{align}\label{eq:czz}
H_{\mathrm{CZZ}} = &\begin{pmatrix}
1 & 0 & 0 & 0 & 0 & 0 & 0 & 0 \\
0 & 1 & 0 & 0 & 0 & 0 & 0 & 0 \\
0 & 0 & 1 & 0 & 0 & 0 & 0 & 0 \\
0 & 0 & 0 & 1 & 0 & 0 & 0 & 0 \\
0 & 0 & 0 & 0 & 1 & 0 & 0 & 0 \\
0 & 0 & 0 & 0 & 0 & -1 & 0 & 0 \\
0 & 0 & 0 & 0 & 0 & 0 & -1 & 0 \\
0 & 0 & 0 & 0 & 0 & 0 & 0& 1
\end{pmatrix}\nonumber\\
&\simeq -\frac{\pi}{4}\left(I_{1} - Z_{1}\right)Z_{2}Z_{3} 
\;=-\frac{\pi}{4}\left( IZZ - ZZZ \right).
\end{align}
The action of the corresponding unitary operator on a computational-basis state is
\[
U_{\text{CZZ}}\ket{q_1q_2q_3} = e^{i\,q_1(-1)^{q_2+q_3}\frac{\pi}{2}} \ket{q_1q_2q_3}.
\]
It is straightforward to verify that when the first qubit (measurement qubit) is in the control state $\ket{1}$,  the odd-parity states of the last two qubits (data qubits), $\ket{101}$ and $\ket{110}$, acquire a nontrivial $+\pi/2$ phase shift,  while the even-parity states, $\ket{100}$ and $\ket{111}$, acquire a $-\pi/2$ phase shift.  This gate serves as a parity-check measurement primitive in the surface code, enabling the readout of $Z$-, $X$-, and $Y$-type stabilizers.  Moreover, the Hamiltonian in Eq.~\eqref{eq:czz} can be implemented via the PCR gate through the decomposition
\begin{align}
U_{\mathrm{CZZ}} 
= S_{1}^{\dagger} H_{3}\, U_{IZX}\!\left(\frac{\pi}{2}\right) \, U_{ZZX}\!\left(-\frac{\pi}{2}\right) \, H_{3},
\end{align}
which requires the corresponding Pauli coefficients to satisfy $\alpha_{ZZX} = -\alpha_{IZX}$,  while all other undesired terms are suppressed.

\subsection{Implementation of $ZZX$}\label{subsec:principle_implementation}
We have demonstrated that utilizing the PCR gate can effectively reduce circuit depth, as both the GHZ state and the $i$Toffoli gate can be implemented in a single shot. We now turn to the realization of the core three-qubit $ZZX$ interaction underlying the PCR gate.

To implement this key mechanism, a minimal superconducting circuit requires only three qubits. In this work, we adopt the coupling strategy used in IBM Quantum’s Eagle processors, which employ a heavy-hexagonal qubit layout. In this architecture, each qubit except those at the edges is coupled to two or three neighbors, forming a lattice reminiscent of tessellated hexagonal edges and vertices.

\begin{figure*}[ht]
    \centering
    \includegraphics[width=0.8\linewidth]{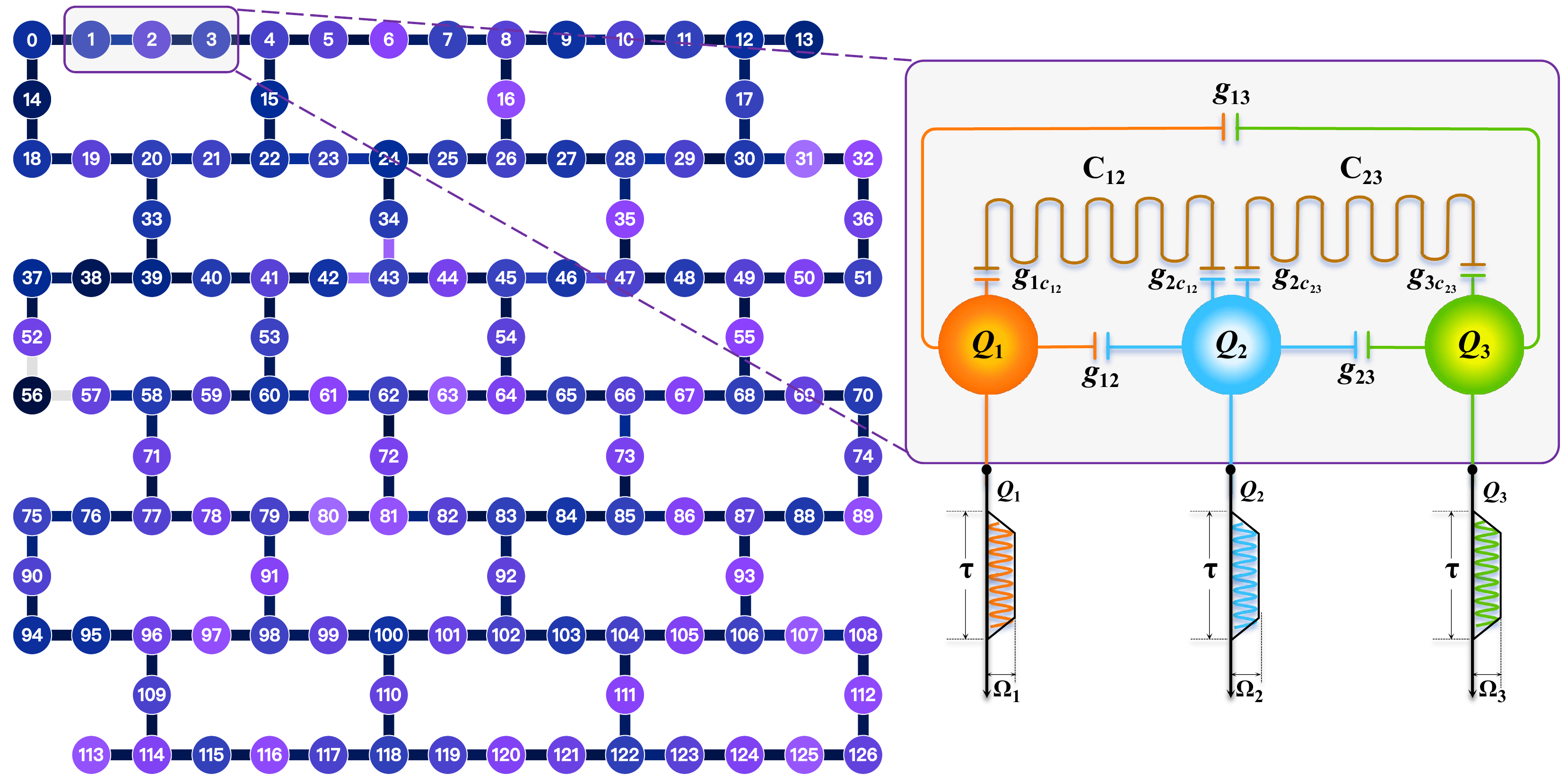}
    \caption{\textbf{Left}: Physical layout of IBM's 127-qubit Eagle quantum processor, \texttt{ibm\_sherbrooke}. \textbf{Right}: Circuit-level representation of a selected three-qubit segment ($Q_1$, $Q_2$, $Q_3$) from the Sherbrooke device.}
    \label{fig:zzx_circuit}
\end{figure*}

We focus on a representative three-qubit unit cell composed of qubits $Q_1$, $Q_2$, and $Q_3$, extracted from the architecture of the \texttt{ibm\_sherbrooke} quantum processor, as illustrated in Fig.~\ref{fig:zzx_circuit}. Within this cell, each pair of neighboring qubits is connected through both a fixed capacitive coupling of strength $g_{ij}$ ($i,j \in \{1,2,3\}, i \neq j$) and a tunable intermediary coupler $C_r$ ($r \in \{12,23\}$) with coupling strengths $g_{iC_r}$ and $g_{jC_r}$. Notably, the terminal qubits $Q_1$ and $Q_3$ do not have dedicated tunable couplers and interact solely via their static capacitive link. 

In the dispersive regime, where $g_{jC_r} \ll |\omega_{C_r} - \omega_{Q_j}|$ with $\omega_{C_r}$ denoting the coupler $C_r$ frequency and $\omega_{Q_j}$ the qubit $Q_j$ frequency, the circuit Hamiltonian can be reduced to the multilevel qubit basis as ($\hbar\equiv 1$)
\begin{align}\label{eq:Hj}
H_{\text{cir}} = & \sum_{Q_j=\{Q_1, Q_2, Q_3\}} \sum_{n_j} \bar{E}_{Q_j}(n_j) |n_j\rangle\langle n_j| \nonumber
\\ & +\sum_{Q_i, Q_j\, (i\neq j)} \, \sum_{m_i, n_j} \sqrt{(m_i+1)(n_j+1)} \nonumber \\
& \times \mathcal{J}_{Q_iQ_j} \left( |m_i, n_j+1\rangle\langle m_i+1, n_j| + \text{H.c.} \right),
\end{align}
where $m_i$ and $n_j$ are two energy levels  in the qubits $Q_i$ and $Q_j$. The quantity $\bar{\omega}_{Q_j}(n_j)$ is defined as the effective transition frequency between levels $n_j$ and $n_j+1$, $\bar{\omega}_{Q_j}(n_j) = \bar{E}_{Q_j}(n_j+1) - \bar{E}_{Q_j}(n_j)$. The term $\mathcal{J}_{Q_iQ_j}$ denotes the set of effective coupling strength between qubit $Q_i$ in state $m_i$ and qubit $Q_j$ in state $n_j$. When either $m_i \geq 1$ or $n_j \geq 1$, we explicitly indicate the excitation levels using index notation: Interactions confined to the computational subspace are denoted as $J_{Q_i Q_j}$, those involving higher excited states as $J_{\overline{Q_i Q_j}}$, and mixed-state interactions as $J_{Q_i\, \overline{Q_j}}$. This notation extends naturally to higher excitation levels; for instance, interactions involving the third excited state are marked with a double overline, such as $J_{\overline{\overline{Q_i}} Q_j}$. For conciseness, we sometimes omit the ``$Q$'' label and refer solely to the qubit indices. Further details are available in Ref.~\cite{xu25surface-code}.

The implementation of the three-qubit interaction follows a protocol analogous to the CR gate. For instance, qubits $Q_1$ and $Q_3$ serve as control qubits, while $Q_2$ is designated as the target. To induce the desired interaction, microwave drives are applied to all three qubits. Each qubit is driven at the dressed frequency of the target qubit $Q_2$, enabling effective control through higher-order interaction pathways. The general form of the drive Hamiltonian is expressed as
\begin{align}
H_{\mathrm{dr}} = \sum_{j=1,2,3} \Omega_{j} \cos\left( \omega_{\mathrm{dr}} t + \phi_{j} \right) \left( \hat{b}_{j} + \hat{b}_{j}^{\dagger} \right),
\end{align}
with $\Omega_{j}$ is the driving amplitude applied on the qubit $Q_j$ with the frequency $\omega_\mathrm{dr}$ and the phase $\phi_{j}$. The operators $\hat{b}_{j}^{\dagger}$ ($\hat{b}_{j}$) is the creation (annihilation) operator for the qubit $Q_j$.

The total Hamiltonian is given by $H_{\text{tot}} = H_{\text{cir}} + H_{\text{dr}}$. To extract the effective interaction relevant for quantum gate implementation, we move to a rotating frame defined by the dressed frequency of the target qubit $Q_2$ and apply the rotating wave approximation (RWA) to eliminate fast-oscillating terms \cite{ansari19superconducting}. In this frame, the Hamiltonian becomes time-independent to leading order and is subsequently block-diagonalized in the computational basis $\{ |n_1, n_3, n_2\rangle \}$, where $n_j\in\{0,\,1\}$ indicates the occupation number of qubit $Q_j$.

Following block diagonalization, the effective Hamiltonian is then rewritten in the Pauli basis as
\begin{align}\label{eq:pauli3}
H_{\text{total}} = \sum_{\substack{A,B \in \{I,Z\} \\ C \in \{I,X,Y,Z\}}} \alpha_{ABC} \, A \otimes B \otimes C.
\end{align}

Here, each coefficient $\alpha_{ABC}$ quantifies the contribution of the corresponding Pauli operator $ABC$ in the effective Hamiltonian. Among these terms, particular attention is devoted to the $ZZX$ component, which represents a genuine three-qubit interaction and constitutes the central focus of our engineering efforts. The objective is to modulate the strength of $\alpha_{ZZX}$ while minimizing the amplitudes of all other undesired components. A perturbative analysis based on the Schrieffer-Wolff transformation (SWT)~\cite{bravyi11schrieffer--wolff}, truncated at first order in the drive amplitude $\Omega$, allows for analytical derivation of these effective interaction terms. The details of this derivation are provided in Appendix~\ref{app:circuit_H}.

In the dispersive regime, the intrinsic strength of the three-body $ZZX$ interaction typically scales as $J^2\Omega/\Delta^2$, rendering it significantly weaker than two-body cross terms like $ZIX$ and $IZX$, which scale more favorably as $J\Omega/\Delta$. This hierarchy reflects the general suppression of higher-order terms predicted by many-body localization theory~\cite{serbyn13local,schreiber15observation,berke22transmon}, where multi-qubit interactions arise perturbatively and thus appear with reduced amplitude. As a result, engineering strong $ZZX$ interactions requires deliberate circuit design strategies that go beyond the standard dispersive approximation. These include: (i) ensuring all three qubits are appreciably coupled, (ii) tuning pairwise detunings into the straddling regime, and (iii) precisely biasing the couplers to access the desired interaction manifold. Recent proposals offer a promising alternative to single-mode tunable couplers: by employing multi-mode coupler architectures, one can significantly enhance effective multi-qubit interactions within more sophisticated circuit layouts~\cite{mcbroom-carroll24entangling}. Table~\ref{tab:comparison} compares the proposed PCR gate with previously demonstrated three-qubit gates based on simultaneous cross-resonance drives. The key distinction is that the proposed method exploits a strong direct three-body $ZZX$ interaction beyond the dispersive regime, whereas the other gates operate in the dispersive regime and rely solely on simultaneous two-body interactions.

\begin{table*}[t]
\centering
\caption{Comparison of the proposed PCR gate with existing three-qubit cross-resonance–based gates.}
\label{tab:comparison}
\begin{tabular}{lccc}
\hline\hline
 & This work (PCR) & Ref.~\cite{kim22high-fidelity} & Ref.~\cite{itoko24three-qubit} \\
\hline
Model Validation & Simulation & Experiment & Experiment\\

Target operation & GHZ/iToffoli/CCNOT/CZZ & iToffoli gate & SCRP gate \\
Topology &$K_3$ (complete graph) &$P_3$ (chain) & $P_3$ (chain)\\
Interaction type & ZZX (ZIX, IZX) &  ZIX, IZX &  ZIX, IZX \\
Operating Regime & Strong/Non-Dispersive & Dispersive& Dispersive \\
Optimization strategy & Search-based algorithm & Berry phase assisted & Echo pulse \\
Stray interaction suppression & Yes & No & Yes \\
Overhead & Static parameter search & Complex pulse sequences & Simultaneous calibration\\
\hline\hline
\end{tabular}
\end{table*}

\section{Gradient Free Optimization Framework}\label{sec:optimization}

To enable high-fidelity three-qubit PCR gate, this section introduces a practical optimization framework designed to selectively amplify the $ZZX$ interaction while suppressing undesired terms. The methodology integrates phase calibration, perturbative initialization, and a gradient-free nonperturbative optimization routine to shape the interaction profile within experimentally accessible constraints.

The procedure begins with calibrating the drive phases $\phi_1$, $\phi_2$, and $\phi_3$ to either $0$ or $\pi$. This phase configuration ensures that all $Y$-type terms in the effective Hamiltonian are eliminated, thereby simplifying the underlying interaction structure and facilitating both analytical tractability and numerical stability.

Following phase calibration, a perturbative analysis is employed to generate an initial estimate for the circuit parameters that favor the emergence of the $ZZX$ term. This initial configuration, derived from analytical expressions valid in the weak-coupling regime, serves as a guided starting point for the subsequent nonperturbative optimization process. Together, these steps establish a robust foundation for efficiently engineering the desired three-body interaction.

Nevertheless, perturbation theory becomes inadequate in regimes where higher-order effects are significant---particularly near the optimal working point, where nonlinear interactions can no longer be ignored. To overcome these limitations and achieve high-fidelity performance, we adopt a \emph{nonperturbative numerical block-diagonalization method}, based on the \emph{principle of least action}~\cite{cederbaum89block,magesan20effective}, which has been modified and implemented in our software package \textbf{CirQubit}~\cite{24cirqubit}. In our simulations, qubits and couplers are modeled with a maximum total excitation number of four, since results are unchanged for higher excitation numbers, see details in Appendix~\ref{app:circuit_H}. This setting is sufficient for the current three-qubit plus two-coupler configurations, and results remain stable under variations in the excitation cutoff.

\begin{figure}[h!]
    \centering
    \includegraphics[width=0.95\linewidth]{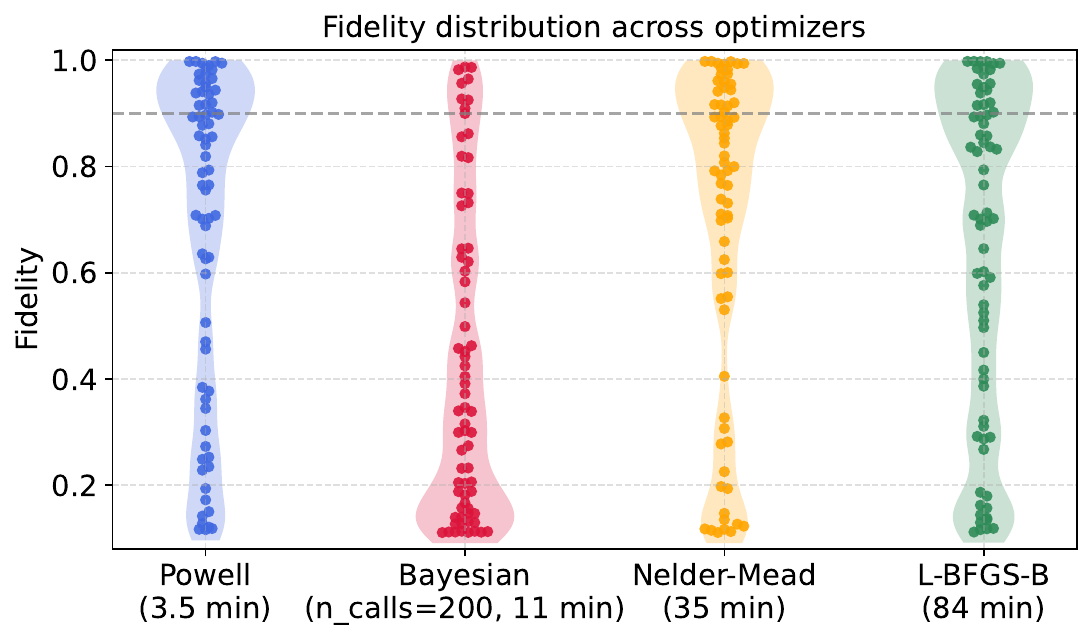}\\
    \includegraphics[width=0.95\linewidth]{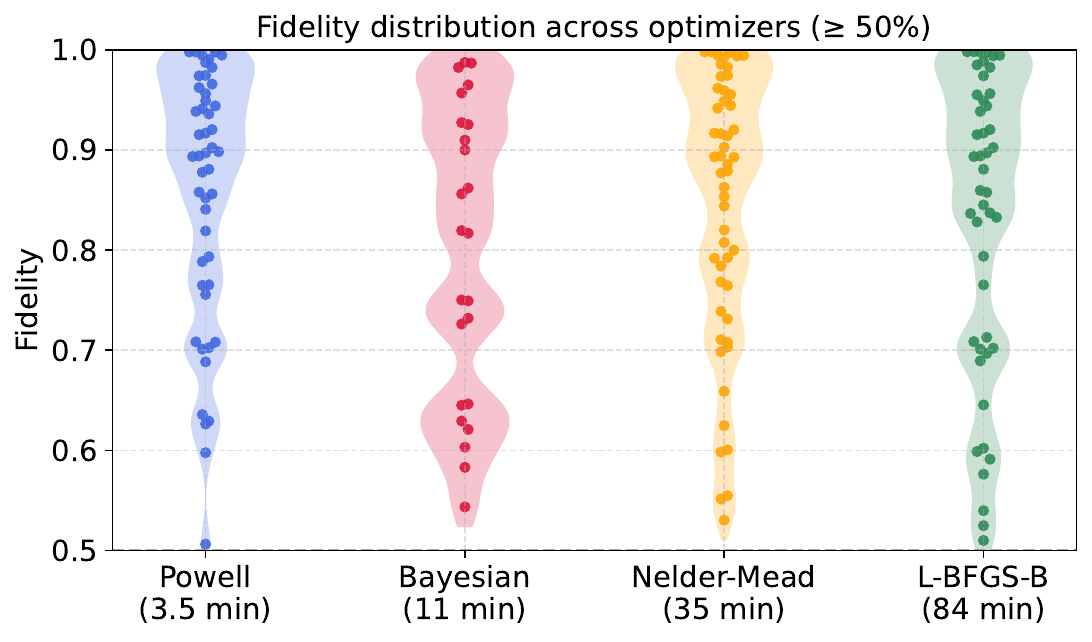}\\
    \includegraphics[width=0.95\linewidth]{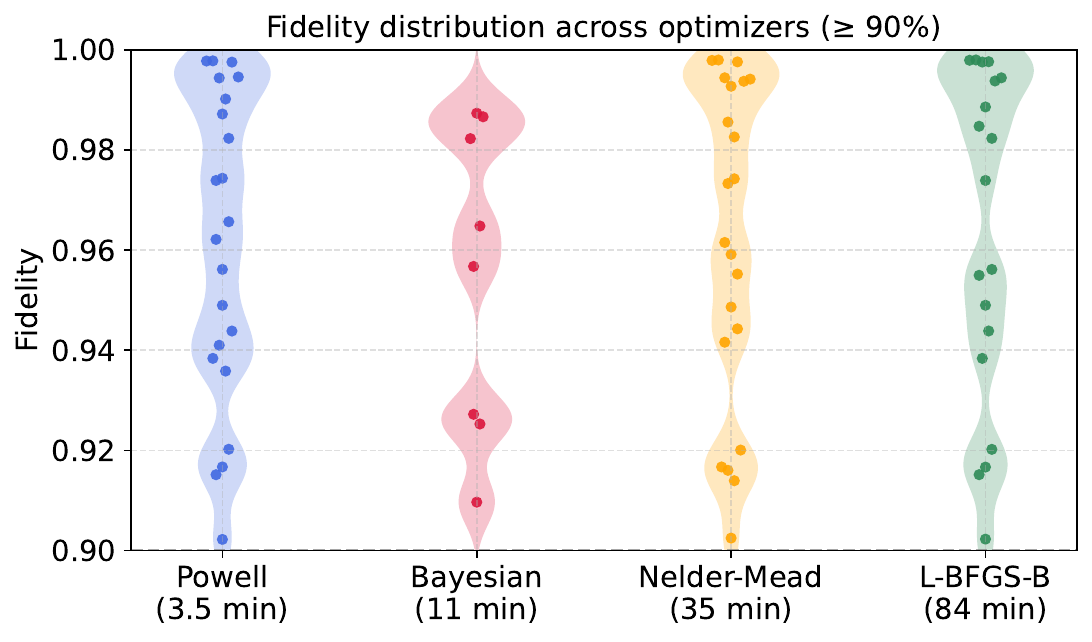}\\
    \includegraphics[width=0.95\linewidth]{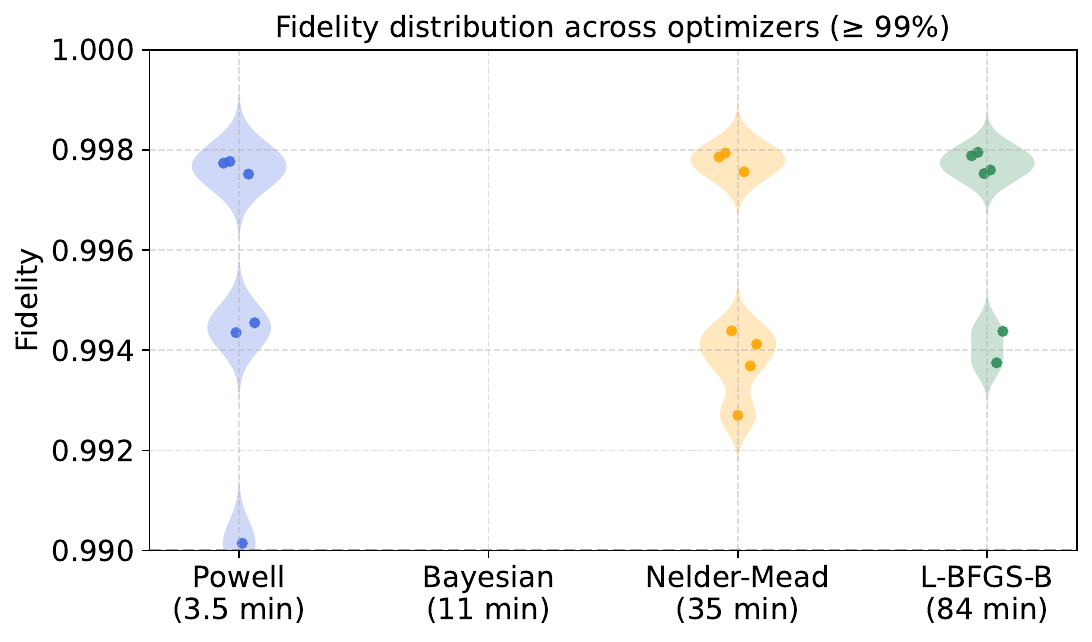}
    \vspace{-0.1in}
 \caption{Comparison of GHZ-state fidelities obtained using different optimization methods: gradient-free Powell, Bayesian optimization, Nelder--Mead, and gradient-based L-BFGS-B. 
The panels display the fidelity distributions for (1) the full range of $F$, (2) $F > 0.5$, (3) $F > 0.9$, and (4) $F > 0.99$. 
All optimization runtimes were benchmarked on an Apple MacBook Pro (2024) equipped with an Apple M4~Pro chip (14-core CPU, 20-core GPU), 24~GB unified memory, and macOS~15.5.
}
    \label{fig:optimizer_com}
\end{figure}

When selecting the optimization strategy for variational parameter tuning, we systematically compare four representative methods: the gradient-free Powell algorithm, Bayesian optimization, the Nelder--Mead simplex method, and the gradient-based L-BFGS-B algorithm. This comparison aims to assess both the convergence efficiency and the accuracy of the resulting quantum states. To provide a quantitative evaluation, we compute the GHZ-state fidelity obtained using the optimized parameters across all unit cells (see details in Sec.~\ref{subsec:ghz_validation}). The numerical results and fidelity distributions corresponding to different fidelity thresholds are summarized in Fig.~\ref{fig:optimizer_com}. All optimization procedures are performed under identical computational conditions on an Apple MacBook Pro (2024) equipped with an Apple M4 Pro chip (14-core CPU, 20-core GPU), 24~GB unified memory, and running macOS~15.5.

Figure~\ref{fig:optimizer_com} presents a comparative analysis of the four optimization methods in terms of GHZ-state fidelity distributions across all unit cells. 
For quantitative reference, the proportions of unit cells achieving fidelities above 50\%, 90\%, and 99\% are as follows: Powell (69.6\%, 31.9\%, 8.7\%), Bayesian optimization with 200 evaluations (34.8\%, 11.6\%, 0.0\%), Nelder--Mead (75.4\%, 31.9\%, 10.1\%), and L-BFGS-B (66.7\%, 27.5\%, 8.7\%). 
Among these, Bayesian optimization exhibits the weakest overall performance, while the gradient-free Powell algorithm achieves a comparable fidelity distribution to Nelder--Mead and L-BFGS-B.

Importantly, the Powell method attains this level of performance with the shortest runtime, averaging approximately 3.5~minutes per optimization, in contrast to the substantially longer runtimes of the Nelder--Mead (35~minutes) and L-BFGS-B (84~minutes) methods. 

Given its favorable trade-off between computational efficiency and accuracy, we adopt the gradient-free conjugate-direction Powell method as our primary optimizer. Its low computational overhead and robustness to stochastic noise make it particularly effective for mitigating the barren plateau problem, especially in near-resonant regimes where gradient-based optimizers often suffer from vanishing signal gradients.

Therefore, the optimization loop employing the Powell algorithm proceeds through the following sequence of steps:

\begin{itemize}
    \item \textbf{Initialization from perturbative estimates}: The process begins with perturbative analysis of a targeted Hamiltonian by estimating circuit parameters such as coupler frequencies and driving amplitude. These estimates provide an informed starting point that guides the optimization toward parameter regimes conducive to strong target interactions.

    \item \textbf{Gradient-free exploration}: A gradient-free heuristic optimization algorithm Powell is used to efficiently navigate the search in the high-dimensional parameter space. The goal is to increase the strength of the desired Pauli term, such as the $ZZX$ interaction, while simultaneously suppressing all unwanted components in the effective Hamiltonian.

    \item \textbf{Nonperturbative Hamiltonian evaluation}: At each step of the optimization, the effective Hamiltonian is reconstructed using the nonperturbative technique. This method accounts for higher-order effects and nonlinearities that are not captured by perturbation theory, ensuring accurate characterization of the system's dynamics~\cite{24cirqubit}.

    \item \textbf{Cost-function-driven refinement}: The Optimization search model iteratively updates the circuit parameters by minimizing a tailored cost function. This function penalizes contributions from undesired Pauli terms and rewards configurations that reinforce the target interaction, thereby guiding the system toward an optimized Hamiltonian structure.
\end{itemize}

This optimization is guided by a carefully designed cost function that balances the enhancement of the wanted terms with the suppression of unwanted interactions:
\begin{align}\label{eq:cost}
\mathcal{L}_{\text{total}} = \mathcal{L}_{\text{wanted}} + \mathcal{L}_{\text{unwanted}} + \mathcal{L}_{\text{constraint}},
\end{align}
where $\mathcal{L}_{\text{wanted}}$ promotes strong, nontrivial desired interactions, $\mathcal{L}_{\text{unwanted}}$ penalizes the presence of undesired Pauli components, and $\mathcal{L}_{\text{constraint}}$ ensures that the circuit parameters remain within a physically valid regime.

Algorithm~\ref{al:powell} outlines the workflow and structure of the Powell search method, with the output corresponding to an optimized set of circuit parameters.

It is worth noting that our search-based approach differs conceptually from quantum optimal control methods, such as GRAPE~\cite{motzoi11optimal}, CRAB~\cite{caneva11chopped}, and GOAT~\cite{machnes18tunable}. Quantum optimal control treats control amplitudes as continuous functions of time and typically employs gradient-based algorithms to maximize gate or state fidelity under physical constraints. In contrast, the present framework focuses on identifying optimal static operating parameters, specifically the tunable coupler frequency and driving amplitude, by exploring the space of experimentally accessible conditions. We regard quantum optimal control as a complementary and natural extension of our approach, which can be applied after suitable operating points are identified to further refine the control protocol and enhance fidelity.

\begin{figure}[h!]
\begin{algorithm}[H]
\caption{Powell Optimization Algorithm}
\label{al:powell}
\begin{algorithmic}[1]
\State \textbf{Input:} Target Pauli term (e.g., $ZZX$), initial circuit parameters $\mathbf{x}_0 \in \mathbb{R}^n$, loss threshold $\varepsilon$, maximum iterations
\State \textbf{Initialization:} Estimate $\mathbf{x}_0$ from perturbative analysis
\State Set direction set: $\{\mathbf{d}_1, \ldots, \mathbf{d}_n\} \gets \{\mathbf{e}_1, \ldots, \mathbf{e}_n\}$ (standard basis)

\While{not converged \textbf{and} iteration count $<$ max iterations}
    \State $\mathbf{x} \gets \mathbf{x}_0$
    \For{$i \gets 1$ \textbf{to} $n$}
        \State Define cost function: $\phi_i(\alpha) \gets \mathcal{L}_{\text{total}}(\mathbf{x} + \alpha \mathbf{d}_i)$
        \State Find optimal step: $\alpha_i^* \gets \arg\min_{\alpha} \phi_i(\alpha)$
        \State Update parameters: $\mathbf{x} \gets \mathbf{x} + \alpha_i^* \mathbf{d}_i$
        \State Evaluate effective Hamiltonian: $\hat{H}_{\text{eff}} \gets \text{CirQubit.NonperturbativeSolver}(\mathbf{x})$
    \EndFor

    \State Compute displacement: $\mathbf{u} \gets \mathbf{x} - \mathbf{x}_0$
    \State Define: $\phi_u(\alpha) \gets \mathcal{L}_{\text{total}}(\mathbf{x}_0 + \alpha \mathbf{u})$
    \State Find: $\alpha^* \gets \arg\min_{\alpha} \phi_u(\alpha)$
    \State Update point: $\mathbf{x}_{\text{new}} \gets \mathbf{x}_0 + \alpha^* \mathbf{u}$
    
    \State Identify longest step: $j \gets \arg\max_{i} \|\alpha_i^* \mathbf{d}_i\|$
    \State Replace direction: $\mathbf{d}_j \gets \mathbf{u}$

    \If{$\|\mathbf{x}_{\text{new}} - \mathbf{x}_0\| < \varepsilon$}
        \State \textbf{break}
    \Else
        \State $\mathbf{x}_0 \gets \mathbf{x}_{\text{new}}$
    \EndIf
\EndWhile

\State \textbf{Output:} Optimized parameters $\mathbf{x}_{\text{new}}$ realizing strong target interaction
\end{algorithmic}
\end{algorithm}
\end{figure}

\section{Validation of the Optimization Algorithm}\label{sec:validation}

To ensure the reliability and effectiveness of the proposed optimization framework, we conduct a detailed validation using two representative applications: the generation of a three-qubit GHZ state and the implementation of an $i$Toffoli gate. These benchmarks illustrate the practical advantages of the optimized PCR gate in realistic settings. For each case, we simulate the time evolution of the system governed by the effective Hamiltonian derived from the optimized circuit parameters. The simulations incorporate each single-qubit gate duration of 30~ns and include decoherence effects modeled using the experimentally measured coherence times of the target quantum processor.

To further validate the method under device-specific conditions, we apply the optimization algorithm to the \texttt{ibm\_sherbrooke} quantum processor, leveraging its scalable architecture to organize qubits into optimized three-qubit unit cells. In particular, we assume a virtual nonzero direct coupling $g_{13}$ to enhance the three-qubit interaction; details are provided in Appendix~\ref{app:circuit_H}. Each unit cell is indexed by $n$ and labeled according to its central qubit, as illustrated in Fig.~\ref{fig:zzx_circuit}.


While IBM's Eagle-generation processors utilize fixed-frequency harmonic resonators to mediate qubit interactions, the newer Heron architecture introduces tunable couplers, offering greater flexibility in engineering interaction strengths. Accordingly, in our simulations, we adopt the following assumptions: qubit frequencies and anharmonicities are held constant, and static coupling strengths are inferred from calibrated $ZZ$ interaction values reported in the processor properties. Under these constraints, the optimization is performed over a reduced set of control variables, namely the drive amplitudes $\Omega_1$, $\Omega_2$, and $\Omega_3$, along with the coupler frequencies $\omega_{C_{12}}$ and $\omega_{C_{23}}$. 

To further simplify the optimization landscape, we fix $\Omega/2\pi=60$ MHz and treat the drive amplitudes of all three qubits in terms of relative scaling factors, defined as $A_1 = \Omega_1 / \Omega$, $A_2 = \Omega_2 / \Omega$ and $A_3 = \Omega_3 / \Omega$. This reparameterization reduces the dimensionality of the search space, facilitating a more efficient and tractable optimization process. The corresponding initialization procedure is detailed in Appendix~\ref{app:algorithm}.

\subsection{Example 1: GHZ state Fidelity}\label{subsec:ghz_validation}

\begin{figure*}[ht]
    \centering
 \includegraphics[width=0.49\linewidth]{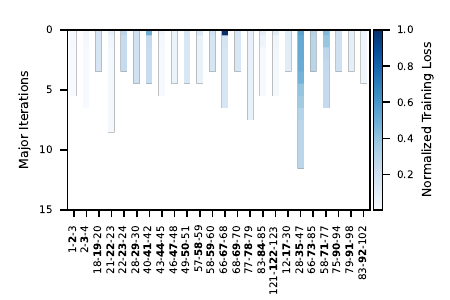}\put(-245,150){\textbf{(a)}}
 \includegraphics[width=.49\linewidth]{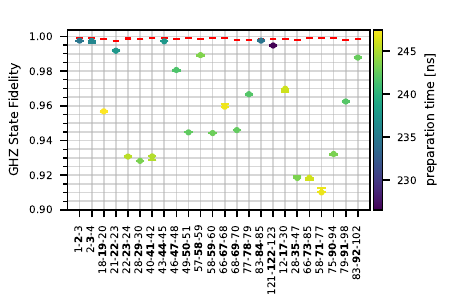}\put(-253,150){\textbf{(b)}}\\
    \vspace{-0.1in}
    \caption{Benchmarking GHZ state preparation on the \texttt{ibm\_sherbrooke} processor. 
(a) Normalized optimization loss versus major iterations, where each iteration updates all parameter search directions once. 
(b) GHZ fidelity for selected three-qubit unit cells using optimized PCR gates ($ZZX_{\pi/2}$) under realistic noise. The colorbar shows evolution time; the red slash marks the coherence limit. Error bars reflect $\pm 2\%$ drive parameter and $\pm 5$ MHz coupler frequency perturbations. $Q_2$ is selected as the \textbf{target} qubit to satisfy the $ZX-$like constraint.}
    \label{fig:ghz_benchmark}
\end{figure*}

To generate a three-qubit GHZ state, the desired interaction is exclusively of the $ZZX$ type, while all other terms are considered unwanted. We evaluate a total of 69 distinct three-qubit circuit configurations, applying the optimization procedure independently to each, using the basis ordering $|Q_1, Q_3, Q_2\rangle$. 
Here, we define the target interaction as $\mathcal{L}_{\text{wanted}} = \mathcal{L}_{ZZX}$, and the resulting training loss across all configurations is summarized in Fig.~\ref{fig:ghz_benchmark}(a).

To assess the fidelity of the generated GHZ state, we first validate the optimization results by substituting the optimized parameters back into the nonperturbative framework to recalculate the effective Pauli coefficients using our software CirQubit~\cite{24cirqubit}, corresponding result is presented in Fig.~\ref{fig:ghz_verify}. These coefficients are subsequently used to reconstruct the full effective Hamiltonian as defined in Eq.~\eqref{eq:pauli3}. We then simulate the time evolution of the initial state $|000\rangle$ under the reconstructed Hamiltonian using corresponding pulse sequences. The PCR gate is implemented using a flat-top Gaussian pulse of duration $\tau$, with 10-ns rise and fall times. Thus, the total implementation time is $t_{\text{GHZ}} = (2 \times 30 + \tau)$ ns, where the two 30-ns single-qubit gates are applied before and after the PCR gate. The fidelity of the resulting state is evaluated by solving the Lindblad master equation. Note that although the optimization is performed under a fixed driving amplitude $\Omega$, the targeted cost function may still hold as long as the Pauli coefficients scale linearly with the drive amplitude~\cite{xu23parasitic-free}. Therefore, the simulation is repeated over a range of $\Omega$ values, and the maximum fidelity is extracted from the evaluated steps. This protocol provides a quantitative assessment of the effectiveness of the $ZZX$-based entangling operation and identifies the optimal parameters for high-fidelity GHZ state preparation.

Figure~\ref{fig:ghz_benchmark}(b) shows that more than one third of the 69 evaluated unit cells achieve GHZ state fidelities above 90\%, with 6 unit cells exceeding 99\% fidelity, demonstrating that the optimized PCR gate consistently enables high-fidelity entanglement. Notably, these high-fidelity states are realized within a total evolution time of 250~ns, underscoring the protocol's robustness and efficiency across diverse hardware configurations. The dominant source of infidelity arises from parasitic stray interactions; however, for unit cells attaining fidelities above 99\%, dynamic $ZZ$ cancellation effectively eliminates such parasitic effects~\cite{xu21zz-freedom}, resulting in a nearly parasitic-free gate~\cite{xu23parasitic-free}.

To assess the protocol’s robustness against experimental imperfections, we introduce random perturbations of $\pm 2\%$ to the optimized drive parameters and $\pm 5$~MHz to the coupler frequencies, with fidelity variations depicted as asymmetric errorbars. The results indicate that only configurations with slightly lower baseline fidelities exhibit sensitivity to these perturbations, while the majority remain largely unaffected. These findings validate the feasibility of implementing native three-qubit gate synthesis on current superconducting quantum hardware.

After generating the GHZ state using the optimized $ZZX$ interaction and corresponding pulse sequence, we evaluate the stability and quality of the resulting entangled state under realistic post-preparation conditions. Once the driving amplitudes are turned off, the system is left to evolve under residual static interactions and decoherence. These parasitic effects, if not properly accounted for, can lead to fidelity degradation~\cite{ku20suppression,zhao20high-contrast,xu21zz-freedom,ni22scalable,ansari0method,ansari0circuit}.

To assess the impact of residual coherent interactions, we consider the static part of the effective Hamiltonian derived from the optimized parameters $
H_{\text{static}} = \alpha_{ZZI} ZZI + \alpha_{ZIZ} ZIZ + \alpha_{IZZ} IZZ + \alpha_{ZZZ} ZZZ$. The time evolution of an ideal GHZ state under this static Hamiltonian $H_{\text{static}}$ leads to the following state after a duration $\tau_p$:
\begin{align}
{\rm GHZ}(\tau_p)\rangle = \frac{e^{-i \alpha_{ZZZ} \tau_p} |000\rangle + e^{i \alpha_{ZZZ} \tau_p} |111\rangle}{\sqrt{2}}.
\end{align}

Interestingly, two-body terms such as $ZZI$, $ZIZ$, and $IZZ$ contribute equally to both $|000\rangle$ and $|111\rangle$, resulting in a global phase that does not alter the quantum state. This is due to the even parity of these operators with respect to the GHZ basis states. In contrast, the $ZZZ$ term introduces a relative phase between the $|000\rangle$ and $|111\rangle$ components due to its odd parity, leading to coherent phase accumulation. Therefore, characterizing and, if necessary, correcting for the $ZZZ$-induced phase shift is essential for maintaining high-fidelity operation in downstream quantum information processing tasks.

One practical approach to mitigate the coherent phase accumulation caused by the $ZZZ$ term is to apply a compensating time-dependent phase correction. This can be achieved, for instance, by implementing a single-qubit $Z$ rotation on any one of the three qubits. Specifically, applying a $Z$ rotation of angle $\alpha_{ZZZ} t$ effectively cancels the relative phase between the $|000\rangle$ and $|111\rangle$ components, restoring the ideal GHZ state. This correction is especially useful when the accumulated phase is deterministic and the interaction strength $\alpha_{ZZZ}$ is well-characterized.

\subsection{Example 2: CCNOT-like Gate Fidelity}\label{subsec:ccnot_validation}

\paragraph*{iToffoli Gate.}  
To enable fast and efficient implementation of the $i$Toffoli gate, we employ a single-shot pulse protocol wherein circuit parameters are optimized to satisfy the symmetry condition $\alpha_{ZZX} = \alpha_{IIX} = -\alpha_{ZIX} = -\alpha_{IZX}$, while simultaneously suppressing all undesired Pauli terms in the effective Hamiltonian. This engineered interaction realizes a native three-qubit gate in the computational basis ordered as $|Q_1, Q_3, Q_2\rangle$, where qubit $Q_2$ acts as the target and $Q_1$, $Q_3$ serve as the controls.

\begin{figure*}[ht]
    \centering
\includegraphics[width=0.49\linewidth]{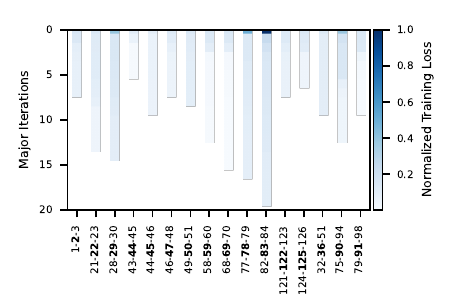}\put(-245,150){\textbf{(a)}}
 \includegraphics[width=.49\linewidth]{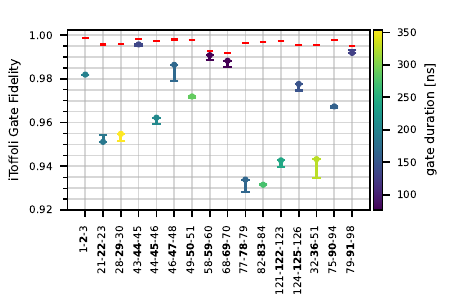}\put(-253,150){\textbf{(b)}}
    \vspace{-0.1in}
\caption{Benchmarking three-qubit $i$Toffoli gate on the \texttt{ibm\_sherbrooke} processor. 
(a) Normalized optimization loss versus major iterations, where each iteration updates all parameter search directions once. 
(b) $i$Toffoli gate fidelity for selected three-qubit unit cells using optimized $ZX$-like interaction under realistic noise. The colorbar shows evolution time; the red slash marks the coherence limit. Error bars reflect $\pm 2\%$ drive parameter and $\pm 5$ MHz coupler frequency perturbations. Qubit $Q_2$ is the \textbf{target}.}

    \label{fig:itoffoli_benchmark}
\end{figure*}

The optimization procedure mirrors that used for GHZ state generation, except that the cost function is replaced with $\mathcal{L}_{i\text{Toffoli}}$, designed specifically to enforce the desired $ZX$-like interaction structure. The optimization loss trajectory is presented in Fig.~\ref{fig:itoffoli_benchmark}(a), and verification of the extracted Pauli coefficients for effective Hamiltonian reconstruction is shown in Fig.~\ref{fig:itoffoli_verify}. The gate is implemented using a flat-top Gaussian pulse envelope, with total duration $t_{i\text{Toffoli}} $~ns.

\paragraph*{CCNOT Gate.}  
The presence of a direct capacitive coupling $g_{13}$ introduces a parasitic $ZZ$ interaction. While this term can impair $i$Toffoli fidelity, it enables an alternative route to implementing a CCNOT gate. In addition to the symmetry conditions required for the $i$Toffoli gate, the CCNOT construction imposes $\alpha_{ZZI} = \alpha_{ZIX}$ and requires calibrated single-qubit rotations on both control qubits, as specified in Eq.~\eqref{eq:ccnot}. The total gate duration is extended to $t_{\text{CCNOT}} = (\tau + 30)$~ns.

\begin{figure*}[ht]
    \centering
\includegraphics[width=0.49\linewidth]{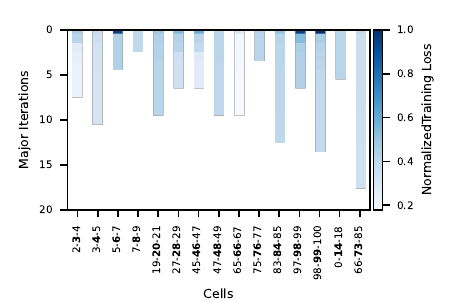}\put(-245,150){\textbf{(a)}}
 \includegraphics[width=0.49\linewidth]{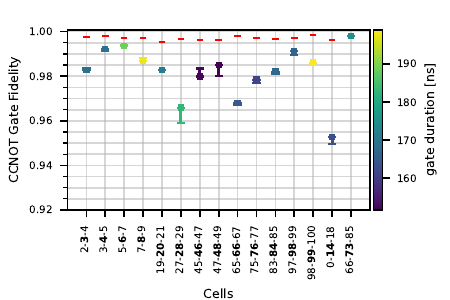}\put(-253,150){\textbf{(b)}}
 \vspace{-0.1in}
\caption{Benchmarking three-qubit CCNOT gate on the \texttt{ibm\_sherbrooke} processor. 
(a) Normalized optimization loss versus major iterations, where each iteration updates all parameter search directions once. 
(b) CCNOT gate fidelity for selected three-qubit unit cells using optimized $ZX$-like interaction under realistic noise. The colorbar shows evolution time; the red slash marks the coherence limit. Error bars reflect $\pm 2\%$ drive parameter and $\pm 5$ MHz coupler frequency perturbations. Qubit $Q_2$ is the \textbf{target}.}
    \label{fig:ccnot_benchmark}
\end{figure*}

To take advantage of this interaction structure, we retrain unit cells that did not achieve high-fidelity $i$Toffoli gates, now targeting CCNOT gate optimization. The corresponding training losses and reconstructed Hamiltonian coefficients are shown in Figs.~\ref{fig:ccnot_benchmark}(a) and \ref{fig:ccnot_verify}, respectively.

Our benchmark spans 69 three-qubit unit cells on the \texttt{ibm\_sherbrooke} architecture. For the \emph{i}Toffoli gate, 16 configurations reach fidelities exceeding $90\%$, with over half completing within $200$~ns---well within superconducting qubit coherence times, as shown in Fig.~\ref{fig:itoffoli_benchmark}(b). In cases where parasitic $ZZ$ interactions degrade \emph{i}Toffoli performance, we repurpose those cells for CCNOT gates. Under an additional symmetry constraint and with post-processed single-qubit rotations, 15 configurations achieve fidelities above $90\%$, as illustrated in Fig.~\ref{fig:ccnot_benchmark}(b). Notably, 3 unit cells for the \emph{i}Toffoli and 4 for the CCNOT exceed $99\%$ fidelity.

In both cases, the gates exhibit robustness to experimental imperfections: random perturbations of $\pm 2\%$ in drive amplitudes and $\pm 5$~MHz in coupler frequencies lead to only marginal fidelity reductions. These results underscore the versatility and resilience of $ZX$-like interaction engineering for realizing high-fidelity, low-latency native multi-qubit gates in superconducting quantum processors.

\subsection{Example 3: CZZ gate}\label{subsec:czz_validation}
To realize a fast and efficient CZZ gate, we adopt a single-shot pulse protocol in which circuit parameters are optimized to satisfy the symmetry condition $\alpha_{ZZX} = -\alpha_{IZX}$, while simultaneously suppressing all other undesired Pauli terms in the effective Hamiltonian.  
The resulting interaction implements a native three-qubit gate in the computational basis ordered as $\ket{Q_1, Q_3, Q_2}$, where $Q_1$ serve as measure qubit, $Q_2$ and $Q_3$ act as the data qubit.  

\begin{figure*}[ht]
    \centering
\includegraphics[width=0.49\linewidth]{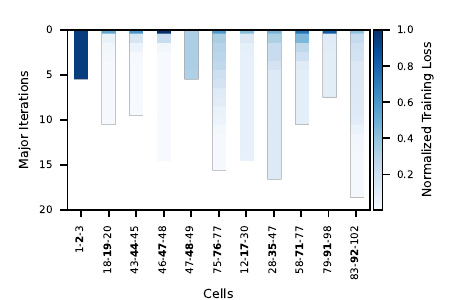}\put(-245,150){\textbf{(a)}}
 \includegraphics[width=0.49\linewidth]{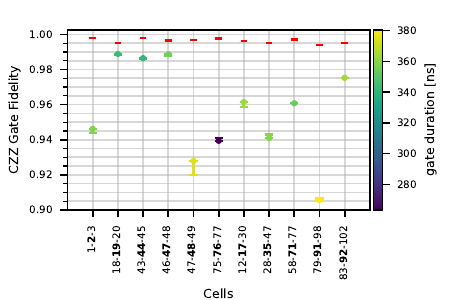}\put(-253,150){\textbf{(b)}}
 \vspace{-0.1in}
\caption{Benchmarking three-qubit CZZ gate on the \texttt{ibm\_sherbrooke} processor. 
(a) Normalized optimization loss versus major iterations, where each iteration updates all parameter search directions once. 
(b) CZZ gate fidelity for selected three-qubit unit cells using optimized $ZZX$ and $IZX$ interaction under realistic noise. The colorbar shows evolution time; the red slash marks the coherence limit. Error bars reflect $\pm 2\%$ drive parameter and $\pm 5$ MHz coupler frequency perturbations. $Q_1$ serves as the measure qubit while $Q_2$ and $Q_3$ act as the data qubits.}
    \label{fig:czz_benchmark}
\end{figure*}

The optimization procedure follows that of the CCNOT-like gate, but with the cost function replaced by $\mathcal{L}_{\text{CZZ}}$, which is tailored to enforce the desired $ZZX$ and $IZX$ interaction structure.  
The optimization loss trajectory is shown in Fig.~\ref{fig:czz_benchmark}(a), while Fig.~\ref{fig:czz_verify} presents the extracted Pauli coefficients from effective Hamiltonian reconstruction. The gate is driven by a flat-top Gaussian pulse envelope with a total duration  $t_{\text{CZZ}} = (\tau + 2\times 30)$ ns, where Hadamard and $S$ gates are applied before and after the PCR gate,  each with a duration of 30 ns. 

Benchmarking results indicate that 11 out of 69 unit cells are capable of implementing the CZZ gate with a fidelity exceeding $90\%$.  The average gate duration is approximately 350 ns, which is slightly longer than that of the CCNOT gate.  The dominant error contributions arise from parasitic interactions, notably $ZZI$ and $ZIZ$.  
Furthermore, random parameter drifts do not significantly affect the performance, demonstrating that the protocol is robust against such variations.

Although this work is primarily based on simulation, the proposed PCR gate is designed using realistic device parameters relevant to current superconducting qubit architectures, without requiring new circuit designs or extreme power levels. Challenges such as microwave calibration, power limitations, and frequency crowding are common to multi-qubit processors~\cite{mohseni25how-to-build, croot25enabling}, but they do not preclude the possibility of implementing the PCR gate experimentally.

The proposed method benefits from increased connectivity, as stronger direct multi-qubit interactions can be realized in modular or highly connected architectures where qubit modules are coupled via common resonators or waveguides. The search-based optimization provides a flexible framework for identifying optimal static operating points and suppressing stray interactions, suggesting that the PCR gate can be extended to larger systems and alternative coupling topologies while maintaining high fidelity.

The dominant sources of error in this work arise from parasitic couplings and decoherence effects. Further performance improvements can be achieved by combining the approach with quantum optimal control techniques, which can effectively suppress leakage to higher energy levels.

\section{Conclusion}
%
%

We have introduced and validated a native three-qubit entangling gate---the Parity Cross-Resonance (PCR) gate---that selectively amplifies the intrinsic $ZZX$ interaction in fixed-frequency superconducting qubit platforms. In contrast to conventional approaches that decompose multi-qubit operations into sequences of two-qubit gates, the PCR gate performs control–control–target and related unitaries in a single coherent step, reducing circuit depth and error accumulation. 

A hybrid optimization framework, combining analytical perturbative modeling with nonperturbative, gradient-free tuning, enables selective enhancement of the desired $ZZX$ term while suppressing spurious couplings. This approach yields robust performance across computational and higher-excitation subspaces, making the gate suitable for diverse applications: GHZ triplet state preparation, Toffoli-class logic, and a native controlled-$ZZ$ (CZZ) operation that directly maps the parity of two data qubits onto a measurement qubit for faster, higher-fidelity stabilizer measurements in surface-code quantum error correction. 

Simulations on selected 69 unit-cell configurations from IBM’s heavy-hex architecture show that multiple subsets of unit cells achieve fidelities above 90\%, several exceed 99\%, and operate within sub-400 ns durations---well within coherence limits. Moreover, the achieved fidelities are not fundamental limits of the architecture. Further improvements are expected for circuits composed exclusively of fully tunable devices when combined with advanced optimal control techniques, which can more effectively suppress crosstalk and mitigate control imperfections.

These results demonstrate that higher-order interactions, often regarded as parasitic, can be repurposed into reliable, scalable primitives. This work lays the foundation for co-designing circuit architectures and control protocols that natively exploit multi-qubit interactions, advancing the scalability and efficiency of next-generation superconducting quantum processors.

\section*{Acknowledgments}
 This research received funding from the Horizon Europe OpenSuperQPlus100 project (Grant Agreement No. 101113946).

 \bibliography{zzx_ml_ref.bib}

@misc{mohseni25how-to-build,
	archiveprefix = {arXiv},
	author = {Masoud Mohseni and Artur Scherer and K. Grace Johnson and Oded Wertheim and Matthew Otten and Navid Anjum Aadit and Yuri Alexeev and Kirk M. Bresniker and Kerem Y. Camsari and Barbara Chapman and Soumitra Chatterjee and Gebremedhin A. Dagnew and Aniello Esposito and Farah Fahim and Marco Fiorentino and Archit Gajjar and Abdullah Khalid and Xiangzhou Kong and Bohdan Kulchytskyy and Elica Kyoseva and Ruoyu Li and P. Aaron Lott and Igor L. Markov and Robert F. McDermott and Giacomo Pedretti and Pooja Rao and Eleanor Rieffel and Allyson Silva and John Sorebo and Panagiotis Spentzouris and Ziv Steiner and Boyan Torosov and Davide Venturelli and Robert J. Visser and Zak Webb and Xin Zhan and Yonatan Cohen and Pooya Ronagh and Alan Ho and Raymond G. Beausoleil and John M. Martinis},
	eprint = {2411.10406},
	title = {How to Build a Quantum Supercomputer: Scaling from Hundreds to Millions of Qubits},
	url = {https://arxiv.org/abs/2411.10406},
	year = {2025}}

@misc{croot25enabling,
	archiveprefix = {arXiv},
	author = {Xanthe Croot and Kasra Nowrouzi and Christopher Spitzer and Carmen G. Almudever and Alexandre Blais and Malcolm Carroll and Jerry Chow and Daniel Friedman and Masao Tokunari and Edoardo Charbon and Vivek Chidambaram and Andrew N. Cleland and David Danovitch and Joseph Emerson and David Gunnarsson and Raymond Laflamme and John Martinis and Robert McDermott and William D. Oliver and Michel Pioro-Ladriere and Yoshiaki Sato and Hidenori Ohata and Kouichi Semba and Irfan Siddiqi},
	eprint = {2512.15001},
	title = {Enabling Technologies for Scalable Superconducting Quantum Computing},
	url = {https://arxiv.org/abs/2512.15001},
	year = {2025}}

@article{old25fault-tolerant,
	author = {Old, Josias and Tasler, Stephan and Hartmann, Michael J. and M\"uller, Markus},
	doi = {10.1103/sblg-fbq4},
	issue = {24},
	journal = {Physical Review Letters},
	month = {Dec},
	numpages = {8},
	pages = {240601},
	publisher = {American Physical Society},
	title = {Fault-Tolerant Stabilizer Measurements in Surface Codes with Three-Qubit Gates},
	url = {https://link.aps.org/doi/10.1103/sblg-fbq4},
	volume = {135},
	year = {2025}}

@article{ansari19superconducting,
	author = {Ansari, M. H.},
	doi = {10.1103/PhysRevB.100.024509},
	issue = {2},
	journal = {Physical Review B},
	month = {Jul},
	numpages = {9},
	pages = {024509},
	publisher = {American Physical Society},
	title = {Superconducting qubits beyond the dispersive regime},
	url = {https://link.aps.org/doi/10.1103/PhysRevB.100.024509},
	volume = {100},
	year = {2019}}

@article{machnes18tunable,
	author = {Machnes, Shai and Ass\'emat, Elie and Tannor, David and Wilhelm, Frank K.},
	doi = {10.1103/PhysRevLett.120.150401},
	issue = {15},
	journal = {Phys. Rev. Lett.},
	month = {Apr},
	numpages = {6},
	pages = {150401},
	publisher = {American Physical Society},
	title = {Tunable, Flexible, and Efficient Optimization of Control Pulses for Practical Qubits},
	url = {https://link.aps.org/doi/10.1103/PhysRevLett.120.150401},
	volume = {120},
	year = {2018}}

@article{motzoi11optimal,
	author = {Motzoi, F. and Gambetta, J. M. and Merkel, S. T. and Wilhelm, F. K.},
	doi = {10.1103/PhysRevA.84.022307},
	issue = {2},
	journal = {Physical Review A},
	month = {Aug},
	numpages = {9},
	pages = {022307},
	publisher = {American Physical Society},
	title = {Optimal control methods for rapidly time-varying Hamiltonians},
	url = {https://link.aps.org/doi/10.1103/PhysRevA.84.022307},
	volume = {84},
	year = {2011}}

@article{caneva11chopped,
	author = {Caneva, Tommaso and Calarco, Tommaso and Montangero, Simone},
	doi = {10.1103/PhysRevA.84.022326},
	issue = {2},
	journal = {Physical Review A},
	month = {Aug},
	numpages = {9},
	pages = {022326},
	publisher = {American Physical Society},
	title = {Chopped random-basis quantum optimization},
	url = {https://link.aps.org/doi/10.1103/PhysRevA.84.022326},
	volume = {84},
	year = {2011}}

@article{nguyen25single-step,
	author = {Nguyen, Minh T.P. and Rimbach-Russ, Maximilian and Vandersypen, Lieven M.K. and Bosco, Stefano},
	doi = {10.1103/kp8s-py9m},
	issue = {3},
	journal = {PRX Quantum},
	month = {Aug},
	numpages = {26},
	pages = {030326},
	publisher = {American Physical Society},
	title = {Single-Step High-Fidelity Three-Qubit Gates by Anisotropic Chiral Interactions},
	url = {https://link.aps.org/doi/10.1103/kp8s-py9m},
	volume = {6},
	year = {2025}}

@article{blumel21power-optimal,
	author = {Bl{\"u}mel, Reinhold and Grzesiak, Nikodem and Pisenti, Neal and Wright, Kenneth and Nam, Yunseong},
	date = {2021/10/08},
	doi = {10.1038/s41534-021-00489-w},
	id = {Bl{\"u}mel2021},
	isbn = {2056-6387},
	journal = {npj Quantum Information},
	number = {1},
	pages = {147},
	title = {Power-optimal, stabilized entangling gate between trapped-ion qubits},
	url = {https://doi.org/10.1038/s41534-021-00489-w},
	volume = {7},
	year = {2021}}

@article{grzesiak20efficient,
	author = {Grzesiak, Nikodem and Bl{\"u}mel, Reinhold and Wright, Kenneth and Beck, Kristin M. and Pisenti, Neal C. and Li, Ming and Chaplin, Vandiver and Amini, Jason M. and Debnath, Shantanu and Chen, Jwo-Sy and Nam, Yunseong},
	date = {2020/06/11},
	doi = {10.1038/s41467-020-16790-9},
	id = {Grzesiak2020},
	isbn = {2041-1723},
	journal = {Nature Communications},
	number = {1},
	pages = {2963},
	title = {Efficient arbitrary simultaneously entangling gates on a trapped-ion quantum computer},
	url = {https://doi.org/10.1038/s41467-020-16790-9},
	volume = {11},
	year = {2020}}

@misc{javadi-abhari25big-cats,
	archiveprefix = {arXiv},
	author = {Ali Javadi-Abhari and Simon Martiel and Alireza Seif and Maika Takita and Ken X. Wei},
	eprint = {2510.09520},
	title = {Big cats: entanglement in 120 qubits and beyond},
	url = {https://arxiv.org/abs/2510.09520},
	year = {2025}}

@article{bruzewicz19dual-species,
	author = {Bruzewicz, C. D. and McConnell, R. and Stuart, J. and Sage, J. M. and Chiaverini, J.},
	date = {2019/11/21},
	doi = {10.1038/s41534-019-0218-z},
	id = {Bruzewicz2019},
	isbn = {2056-6387},
	journal = {npj Quantum Information},
	number = {1},
	pages = {102},
	title = {Dual-species, multi-qubit logic primitives for $\mathrm{Ca}^{+}/\mathrm{Sr}^{+}$ trapped-ion crystals},
	url = {https://doi.org/10.1038/s41534-019-0218-z},
	volume = {5},
	year = {2019}}

@article{hahn19integrated,
	author = {Hahn, H. and Zarantonello, G. and Schulte, M. and Bautista-Salvador, A. and Hammerer, K. and Ospelkaus, C.},
	date = {2019/08/16},
	doi = {10.1038/s41534-019-0184-5},
	id = {Hahn2019},
	isbn = {2056-6387},
	journal = {npj Quantum Information},
	number = {1},
	pages = {70},
	title = {Integrated ${}^{9}\mathrm{Be}^{+}$ multi-qubit gate device for the ion-trap quantum computer},
	url = {https://doi.org/10.1038/s41534-019-0184-5},
	volume = {5},
	year = {2019}}

@article{cao24multi-qubit,
	author = {Cao, Alec and Eckner, William J. and Lukin Yelin, Theodor and Young, Aaron W. and Jandura, Sven and Yan, Lingfeng and Kim, Kyungtae and Pupillo, Guido and Ye, Jun and Darkwah Oppong, Nelson and Kaufman, Adam M.},
	date = {2024/10/01},
	doi = {10.1038/s41586-024-07913-z},
	id = {Cao2024},
	isbn = {1476-4687},
	journal = {Nature},
	number = {8033},
	pages = {315--320},
	title = {Multi-qubit gates and Schr{\"o}dinger cat states in an optical clock},
	url = {https://doi.org/10.1038/s41586-024-07913-z},
	volume = {634},
	year = {2024}}

@article{khazali20fast,
	author = {Khazali, Mohammadsadegh and M\o{}lmer, Klaus},
	doi = {10.1103/PhysRevX.10.021054},
	issue = {2},
	journal = {Physical Review X},
	month = {Jun},
	numpages = {16},
	pages = {021054},
	publisher = {American Physical Society},
	title = {Fast Multiqubit Gates by Adiabatic Evolution in Interacting Excited-State Manifolds of Rydberg Atoms and Superconducting Circuits},
	url = {https://link.aps.org/doi/10.1103/PhysRevX.10.021054},
	volume = {10},
	year = {2020}}

@article{evered23high-fidelity,
	author = {Evered, Simon J. and Bluvstein, Dolev and Kalinowski, Marcin and Ebadi, Sepehr and Manovitz, Tom and Zhou, Hengyun and Li, Sophie H. and Geim, Alexandra A. and Wang, Tout T. and Maskara, Nishad and Levine, Harry and Semeghini, Giulia and Greiner, Markus and Vuleti{\'c}, Vladan and Lukin, Mikhail D.},
	date = {2023/10/01},
	doi = {10.1038/s41586-023-06481-y},
	id = {Evered2023},
	isbn = {1476-4687},
	journal = {Nature},
	number = {7982},
	pages = {268--272},
	title = {High-fidelity parallel entangling gates on a neutral-atom quantum computer},
	url = {https://doi.org/10.1038/s41586-023-06481-y},
	volume = {622},
	year = {2023}}

@article{levine19parallel,
	author = {Levine, Harry and Keesling, Alexander and Semeghini, Giulia and Omran, Ahmed and Wang, Tout T. and Ebadi, Sepehr and Bernien, Hannes and Greiner, Markus and Vuleti\ifmmode \acute{c}\else \'{c}\fi{}, Vladan and Pichler, Hannes and Lukin, Mikhail D.},
	doi = {10.1103/PhysRevLett.123.170503},
	issue = {17},
	journal = {Phys. Rev. Lett.},
	month = {Oct},
	numpages = {6},
	pages = {170503},
	publisher = {American Physical Society},
	title = {Parallel Implementation of High-Fidelity Multiqubit Gates with Neutral Atoms},
	url = {https://link.aps.org/doi/10.1103/PhysRevLett.123.170503},
	volume = {123},
	year = {2019}}

@article{chen25efficient,
	author = {Chen, Zhen and Liu, Weiyang and Ma, Yanjun and Sun, Weijie and Wang, Ruixia and Wang, He and Xu, Huikai and Xue, Guangming and Yan, Haisheng and Yang, Zhen and Ding, Jiayu and Gao, Yang and Li, Feiyu and Zhang, Yujia and Zhang, Zikang and Jin, Yirong and Yu, Haifeng and Chen, Jianxin and Yan, Fei},
	date = {2025/09/01},
	doi = {10.1038/s41567-025-02990-x},
	id = {Chen2025},
	isbn = {1745-2481},
	journal = {Nature Physics},
	number = {9},
	pages = {1489--1496},
	title = {Efficient implementation of arbitrary two-qubit gates using unified control},
	url = {https://doi.org/10.1038/s41567-025-02990-x},
	volume = {21},
	year = {2025}}

@article{yang04efficient,
	author = {Yang, Chui-Ping and Chu, Shih-I and Han, Siyuan},
	doi = {10.1103/PhysRevA.70.022329},
	issue = {2},
	journal = {Physical Review A},
	month = {Aug},
	numpages = {8},
	pages = {022329},
	publisher = {American Physical Society},
	title = {Efficient many-party controlled teleportation of multiqubit quantum information via entanglement},
	url = {https://link.aps.org/doi/10.1103/PhysRevA.70.022329},
	volume = {70},
	year = {2004}}

@article{bose98multiparticle,
	author = {Bose, S. and Vedral, V. and Knight, P. L.},
	doi = {10.1103/PhysRevA.57.822},
	issue = {2},
	journal = {Physical Review A},
	month = {Feb},
	numpages = {0},
	pages = {822--829},
	publisher = {American Physical Society},
	title = {Multiparticle generalization of entanglement swapping},
	url = {https://link.aps.org/doi/10.1103/PhysRevA.57.822},
	volume = {57},
	year = {1998}}

@article{buhrman10nonlocality,
	author = {Buhrman, Harry and Cleve, Richard and Massar, Serge and de Wolf, Ronald},
	doi = {10.1103/RevModPhys.82.665},
	issue = {1},
	journal = {Reviews of Modern Physics},
	month = {Mar},
	numpages = {0},
	pages = {665--698},
	publisher = {American Physical Society},
	title = {Nonlocality and communication complexity},
	url = {https://link.aps.org/doi/10.1103/RevModPhys.82.665},
	volume = {82},
	year = {2010}}

@article{preskill98reliable,
	author = {Preskill, John},
	doi = {10.1098/rspa.1998.0167},
	journal = {Proceedings of the Royal Society of London. Series A: Mathematical, Physical and Engineering Sciences},
	number = {1969},
	pages = {385-410},
	title = {Reliable quantum computers},
	url = {https://royalsocietypublishing.org/doi/abs/10.1098/rspa.1998.0167},
	volume = {454},
	year = {1998}}

@article{divincenzo96fault-tolerant,
	author = {DiVincenzo, David P. and Shor , Peter W.},
	doi = {10.1103/PhysRevLett.77.3260},
	issue = {15},
	journal = {Physical Review Letters},
	month = {Oct},
	numpages = {0},
	pages = {3260--3263},
	publisher = {American Physical Society},
	title = {Fault-Tolerant Error Correction with Efficient Quantum Codes},
	url = {https://link.aps.org/doi/10.1103/PhysRevLett.77.3260},
	volume = {77},
	year = {1996}}

@article{hillery99quantum,
	author = {Hillery, Mark and Bu\ifmmode \check{z}\else \v{z}\fi{}ek, Vladim\'{\i}r and Berthiaume, Andr\'e},
	doi = {10.1103/PhysRevA.59.1829},
	issue = {3},
	journal = {Physical Review A},
	month = {Mar},
	numpages = {0},
	pages = {1829--1834},
	publisher = {American Physical Society},
	title = {Quantum secret sharing},
	url = {https://link.aps.org/doi/10.1103/PhysRevA.59.1829},
	volume = {59},
	year = {1999}}

@misc{xu25surface-code,
	archiveprefix = {arXiv},
	author = {Xuexin Xu and Kuljeet Kaur and Chlo{\'e} Vignes and Mohammad H. Ansari and John M. Martinis},
	eprint = {2507.06201},
	title = {Surface-Code Hardware Hamiltonian},
	url = {https://arxiv.org/abs/2507.06201},
	year = {2025}}

@article{itoko24three-qubit,
	author = {Itoko, Toshinari and Malekakhlagh, Moein and Kanazawa, Naoki and Takita, Maika},
	doi = {10.1103/PhysRevApplied.21.034018},
	issue = {3},
	journal = {Physical Review Applied},
	month = {Mar},
	numpages = {14},
	pages = {034018},
	publisher = {American Physical Society},
	title = {Three-qubit parity gate via simultaneous cross-resonance drives},
	url = {https://link.aps.org/doi/10.1103/PhysRevApplied.21.034018},
	volume = {21},
	year = {2024}}

@misc{tasler25optimizing,
	archiveprefix = {arXiv},
	author = {Stephan Tasler and Josias Old and Lukas Heunisch and Verena Feulner and Timo Eckstein and Markus M{\"u}ller and Michael J. Hartmann},
	eprint = {2506.09028},
	title = {Optimizing Superconducting Three-Qubit Gates for Surface-Code Error Correction},
	url = {https://arxiv.org/abs/2506.09028},
	year = {2025}}

@article{cho19emergence,
	author = {Cho, Young-Wook and Kim, Yosep and Choi, Yeon-Ho and Kim, Yong-Su and Han, Sang-Wook and Lee, Sang-Yun and Moon, Sung and Kim, Yoon-Ho},
	date = {2019/07/01},
	doi = {10.1038/s41567-019-0482-z},
	id = {Cho2019},
	isbn = {1745-2481},
	journal = {Nature Physics},
	number = {7},
	pages = {665--670},
	title = {Emergence of the geometric phase from quantum measurement back-action},
	url = {https://doi.org/10.1038/s41567-019-0482-z},
	volume = {15},
	year = {2019}}

@misc{shende08on-the-cnot-cost,
	archiveprefix = {arXiv},
	author = {Vivek V. Shende and Igor L. Markov},
	eprint = {0803.2316},
	title = {On the CNOT-cost of TOFFOLI gates},
	url = {https://arxiv.org/abs/0803.2316},
	year = {2008}}

@article{pazem25error,
	author = {Pazem, Jos{\'e}phine and Ansari, Mohammad H},
	journal = {Scientific Reports},
	number = {1},
	pages = {2257},
	publisher = {Nature Publishing Group UK London},
	title = {Error mitigation in brainbox quantum autoencoders},
	url = {https://doi.org/10.1038/s41598-024-84171-z},
	volume = {15},
	year = {2025}}

@article{mcbroom-carroll24entangling,
	author = {McBroom-Carroll, T. and Schlabes, A. and Xu, X. and Ku, J. and Cole, B. and Indrajeet, S. and LaHaye, M. D. and Ansari, M. H. and Plourde, B. L. T.},
	doi = {10.1103/PRXQuantum.5.020325},
	issue = {2},
	journal = {PRX Quantum},
	month = {May},
	numpages = {17},
	pages = {020325},
	publisher = {American Physical Society},
	title = {Entangling Interactions Between Artificial Atoms Mediated by a Multimode Left-Handed Superconducting Ring Resonator},
	url = {https://link.aps.org/doi/10.1103/PRXQuantum.5.020325},
	volume = {5},
	year = {2024}}

@article{ansari0circuit,
	author = {Mohammed Ansari and Xuexin Xu},
	journal = {Germany Patent DE102020201688B3},
	title = {Circuit with coupled qubits with different anharmonic energy spectrum},
	url = {https://patents.google.com/patent/DE102020201688B3/en?inventor=xuexin+xu&oq=xuexin+xu},
	year = {issued on Jul. 29, 2021}}

@article{ansari0method,
	author = {Mohammed Ansari and Xuexin Xu},
	journal = {European Patent EP4205044},
	title = {Method of operating a circuit with a first and a second qubit},
	url = {https://register.epo.org/application?number=EP21763382&tab=main},
	year = {issued on May 07, 2023}}

@article{ansari0three,
	author = {Mohammed Ansari and Xuexin Xu},
	journal = {European Patent EP4571596},
	title = {THREE-QUBIT GATE AND METHOD FOR ITS REALIZATION},
	url = {https://register.epo.org/application?number=EP23215766&tab=main},
	year = {Issued on Jun. 18, 2025}}

@article{ku20suppression,
	author = {Ku, Jaseung and Xu, Xuexin and Brink, Markus and McKay, David C. and Hertzberg, Jared B. and Ansari, Mohammad H. and Plourde, B. L. T.},
	doi = {10.1103/PhysRevLett.125.200504},
	issue = {20},
	journal = {Physical Review Letters},
	month = {Nov},
	numpages = {6},
	pages = {200504},
	publisher = {American Physical Society},
	title = {Suppression of Unwanted $ZZ$ Interactions in a Hybrid Two-Qubit System},
	url = {https://link.aps.org/doi/10.1103/PhysRevLett.125.200504},
	volume = {125},
	year = {2020}}

@article{xu21zz-freedom,
	author = {Xu, Xuexin and Ansari, M.H.},
	doi = {10.1103/PhysRevApplied.15.064074},
	issue = {6},
	journal = {Physical Review Applied},
	month = {Jun},
	numpages = {14},
	pages = {064074},
	publisher = {American Physical Society},
	title = {$ZZ$ Freedom in Two-Qubit Gates},
	url = {https://link.aps.org/doi/10.1103/PhysRevApplied.15.064074},
	volume = {15},
	year = {2021}}

@article{xu23parasitic-free,
	author = {Xu, Xuexin and Ansari, M.},
	doi = {10.1103/PhysRevApplied.19.024057},
	issue = {2},
	journal = {Physical Review Applied},
	month = {Feb},
	numpages = {14},
	pages = {024057},
	publisher = {American Physical Society},
	title = {Parasitic-Free Gate: An Error-Protected Cross-Resonance Switch in Weakly Tunable Architectures},
	url = {https://link.aps.org/doi/10.1103/PhysRevApplied.19.024057},
	volume = {19},
	year = {2023}}

@article{zhao20high-contrast,
	author = {Zhao, Peng and Xu, Peng and Lan, Dong and Chu, Ji and Tan, Xinsheng and Yu, Haifeng and Yu, Yang},
	doi = {10.1103/PhysRevLett.125.200503},
	issue = {20},
	journal = {Physical Review Letters},
	month = {Nov},
	numpages = {6},
	pages = {200503},
	publisher = {American Physical Society},
	title = {High-Contrast $ZZ$ Interaction Using Superconducting Qubits with Opposite-Sign Anharmonicity},
	url = {https://link.aps.org/doi/10.1103/PhysRevLett.125.200503},
	volume = {125},
	year = {2020}}

@article{corcoles13process,
	author = {C\'orcoles, A. D. and Gambetta, Jay M. and Chow, Jerry M. and Smolin, John A. and Ware, Matthew and Strand, Joel and Plourde, B. L. T. and Steffen, M.},
	doi = {10.1103/PhysRevA.87.030301},
	issue = {3},
	journal = {Physical Review A},
	month = {Mar},
	numpages = {4},
	pages = {030301},
	publisher = {American Physical Society},
	title = {Process verification of two-qubit quantum gates by randomized benchmarking},
	url = {https://link.aps.org/doi/10.1103/PhysRevA.87.030301},
	volume = {87},
	year = {2013}}

@article{rigetti10fully,
	author = {Rigetti, Chad and Devoret, Michel},
	doi = {10.1103/PhysRevB.81.134507},
	issue = {13},
	journal = {Physical Review B},
	month = {Apr},
	numpages = {7},
	pages = {134507},
	publisher = {American Physical Society},
	title = {Fully microwave-tunable universal gates in superconducting qubits with linear couplings and fixed transition frequencies},
	url = {https://link.aps.org/doi/10.1103/PhysRevB.81.134507},
	volume = {81},
	year = {2010}}

@article{sung21realization,
	author = {Sung, Youngkyu and Ding, Leon and Braum\"uller, Jochen and Veps\"al\"ainen, Antti and Kannan, Bharath and Kjaergaard, Morten and Greene, Ami and Samach, Gabriel O. and McNally, Chris and Kim, David and Melville, Alexander and Niedzielski, Bethany M. and Schwartz, Mollie E. and Yoder, Jonilyn L. and Orlando, Terry P. and Gustavsson, Simon and Oliver, William D.},
	doi = {10.1103/PhysRevX.11.021058},
	issue = {2},
	journal = {Physical Review X},
	month = {Jun},
	numpages = {32},
	pages = {021058},
	publisher = {American Physical Society},
	title = {Realization of High-Fidelity CZ and $ZZ$-Free iSWAP Gates with a Tunable Coupler},
	url = {https://link.aps.org/doi/10.1103/PhysRevX.11.021058},
	volume = {11},
	year = {2021}}

@article{menke22demonstration,
	author = {Menke, Tim and Banner, William P. and Bergamaschi, Thomas R. and Di Paolo, Agustin and Veps\"al\"ainen, Antti and Weber, Steven J. and Winik, Roni and Melville, Alexander and Niedzielski, Bethany M. and Rosenberg, Danna and Serniak, Kyle and Schwartz, Mollie E. and Yoder, Jonilyn L. and Aspuru-Guzik, Al\'an and Gustavsson, Simon and Grover, Jeffrey A. and Hirjibehedin, Cyrus F. and Kerman, Andrew J. and Oliver, William D.},
	doi = {10.1103/PhysRevLett.129.220501},
	issue = {22},
	journal = {Physical Review Letters},
	month = {Nov},
	numpages = {6},
	pages = {220501},
	publisher = {American Physical Society},
	title = {Demonstration of Tunable Three-Body Interactions between Superconducting Qubits},
	url = {https://link.aps.org/doi/10.1103/PhysRevLett.129.220501},
	volume = {129},
	year = {2022}}

@article{xu24lattice,
	author = {Xu, Xuexin and Manabputra and Vignes, Chlo\'e and Ansari, Mohammad H. and Martinis, John M.},
	doi = {10.1103/PhysRevApplied.22.064030},
	issue = {6},
	journal = {Physical Review Applied},
	month = {Dec},
	numpages = {20},
	pages = {064030},
	publisher = {American Physical Society},
	title = {Lattice Hamiltonians and stray interactions within quantum processors},
	url = {https://link.aps.org/doi/10.1103/PhysRevApplied.22.064030},
	volume = {22},
	year = {2024}}

@misc{24cirqubit,
	howpublished = {Software suite for simulating superconducting qubits,},
	lastchecked = {CirQubit cirqubit.com},
	title = {CirQubit},
	url = {https://cirqubit.com},
	year = {2024}}

@article{chow12universal,
	author = {Chow, Jerry M. and Gambetta, Jay M. and C\'{o}rcoles, A. D. and Merkel, Seth T. and Smolin, John A. and Rigetti, Chad and Poletto, S. and Keefe, George A. and Rothwell, Mary B. and Rozen, J. R. and Ketchen, Mark B. and Steffen, M.},
	doi = {10.1103/PhysRevLett.109.060501},
	issue = {6},
	journal = {Physical Review Letters},
	month = {Aug},
	numpages = {5},
	pages = {060501},
	publisher = {American Physical Society},
	title = {Universal Quantum Gate Set Approaching Fault-Tolerant Thresholds with Superconducting Qubits},
	url = {https://link.aps.org/doi/10.1103/PhysRevLett.109.060501},
	volume = {109},
	year = {2012}}

@article{chow11simple,
	author = {Chow, Jerry M. and C\'orcoles, A. D. and Gambetta, Jay M. and Rigetti, Chad and Johnson, B. R. and Smolin, John A. and Rozen, J. R. and Keefe, George A. and Rothwell, Mary B. and Ketchen, Mark B. and Steffen, M.},
	doi = {10.1103/physrevlett.107.080502},
	journal = {Physical Review Letters},
	number = {8},
	pages = {080502},
	publisher = {APS},
	title = {Simple all-microwave entangling gate for fixed-frequency superconducting qubits},
	volume = {107},
	year = {2011}}

@article{sheldon16procedure,
	author = {Sheldon, Sarah and Magesan, Easwar and Chow, Jerry M. and Gambetta, Jay M.},
	doi = {10.1103/PhysRevA.93.060302},
	issue = {6},
	journal = {Physical Review A},
	month = {Jun},
	numpages = {5},
	pages = {060302(R)},
	title = {Procedure for systematically tuning up cross-talk in the cross-resonance gate},
	url = {https://link.aps.org/doi/10.1103/PhysRevA.93.060302},
	volume = {93},
	year = {2016}}

@article{magesan20effective,
	author = {Magesan, Easwar and Gambetta, Jay M.},
	doi = {10.1103/PhysRevA.101.052308},
	issue = {5},
	journal = {Physical Review A},
	month = {May},
	numpages = {15},
	pages = {052308},
	publisher = {American Physical Society},
	title = {Effective Hamiltonian models of the cross-resonance gate},
	url = {https://link.aps.org/doi/10.1103/PhysRevA.101.052308},
	volume = {101},
	year = {2020}}

@book{nielsen02quantum,
	author = {Nielsen, Michael A and Chuang, Isaac},
	doi = {10.1017/cbo9780511976667},
	publisher = {American Association of Physics Teachers},
	title = {Quantum computation and quantum information},
	year = {2002}}

@article{schreiber15observation,
	author = {Michael Schreiber and Sean S. Hodgman and Pranjal Bordia and Henrik P. L{\"u}schen and Mark H. Fischer and Ronen Vosk and Ehud Altman and Ulrich Schneider and Immanuel Bloch},
	doi = {10.1126/science.aaa7432},
	journal = {Science},
	number = {6250},
	pages = {842-845},
	title = {Observation of many-body localization of interacting fermions in a quasirandom optical lattice},
	volume = {349},
	year = {2015}}

@article{serbyn13local,
	author = {Serbyn, Maksym and Papi\ifmmode \acute{c}\else \'{c}\fi{}, Z. and Abanin, Dmitry A.},
	doi = {10.1103/PhysRevLett.111.127201},
	issue = {12},
	journal = {Physical Review Letters},
	month = {Sep},
	numpages = {5},
	pages = {127201},
	publisher = {American Physical Society},
	title = {Local Conservation Laws and the Structure of the Many-Body Localized States},
	url = {https://link.aps.org/doi/10.1103/PhysRevLett.111.127201},
	volume = {111},
	year = {2013}}

@article{warren23extensive,
	author = {Warren, Christopher W. and Fern{\'a}ndez-Pend{\'a}s, Jorge and Ahmed, Shahnawaz and Abad, Tahereh and Bengtsson, Andreas and Bizn{\'a}rov{\'a}, Janka and Debnath, Kamanasish and Gu, Xiu and Kri{\v z}an, Christian and Osman, Amr and Fadavi Roudsari, Anita and Delsing, Per and Johansson, G{\"o}ran and Frisk Kockum, Anton and Tancredi, Giovanna and Bylander, Jonas},
	doi = {10.1038/s41534-023-00711-x},
	journal = {npj Quantum Information},
	number = {1},
	pages = {44},
	publisher = {Nature Publishing Group UK London},
	title = {Extensive characterization and implementation of a family of three-qubit gates at the coherence limit},
	volume = {9},
	year = {2023}}

@article{ni22scalable,
	author = {Ni, Zhongchu and Li, Sai and Zhang, Libo and Chu, Ji and Niu, Jingjing and Yan, Tongxing and Deng, Xiuhao and Hu, Ling and Li, Jian and Zhong, Youpeng and Liu, Song and Yan, Fei and Xu, Yuan and Yu, Dapeng},
	doi = {https://doi.org/10.1103/PhysRevLett.129.040502},
	journal = {Physical Review Letters},
	number = {4},
	pages = {040502},
	publisher = {APS},
	title = {Scalable Method for Eliminating Residual ZZ Interaction between Superconducting Qubits},
	volume = {129},
	year = {2022}}

@article{gu21fast,
	author = {Gu, Xiu and Fern{\'a}ndez-Pend{\'a}s, Jorge and Vikst{\aa}l, Pontus and Abad, Tahereh and Warren, Christopher and Bengtsson, Andreas and Tancredi, Giovanna and Shumeiko, Vitaly and Bylander, Jonas and Johansson, G{\"o}ran and Anton Frisk Kockum},
	doi = {10.1103/prxquantum.2.040348},
	journal = {PRX Quantum},
	number = {4},
	pages = {040348},
	publisher = {APS},
	title = {Fast multiqubit gates through simultaneous two-qubit gates},
	volume = {2},
	year = {2021}}

@article{kim22high-fidelity,
	author = {Kim, Yosep and Morvan, Alexis and Nguyen, Long B. and Naik, Ravi K. and J{\"u}nger, Christian and Chen, Larry and Kreikebaum, John Mark and Santiago, David I. and Siddiqi, Irfan},
	date = {2022/07/01},
	doi = {10.1038/s41567-022-01590-3},
	id = {Kim2022},
	isbn = {1745-2481},
	journal = {Nature Physics},
	number = {7},
	pages = {783--788},
	title = {High-fidelity three-qubit iToffoli gate for fixed-frequency superconducting qubits},
	url = {https://doi.org/10.1038/s41567-022-01590-3},
	volume = {18},
	year = {2022}}

@article{kandala21demonstration,
	author = {Kandala, A. and Wei, K. X. and Srinivasan, S. and Magesan, E. and Carnevale, S. and Keefe, G. A. and Klaus, D. and Dial, O. and McKay, D. C.},
	doi = {10.1103/PhysRevLett.127.130501},
	issue = {13},
	journal = {Physical Review Letters},
	month = {Sep},
	numpages = {6},
	pages = {130501},
	publisher = {American Physical Society},
	title = {Demonstration of a High-Fidelity cnot Gate for Fixed-Frequency Transmons with Engineered $ZZ$ Suppression},
	url = {https://link.aps.org/doi/10.1103/PhysRevLett.127.130501},
	volume = {127},
	year = {2021}}

@article{bravyi11schrieffer--wolff,
	author = {Sergey Bravyi and David P. DiVincenzo and Daniel Loss},
	doi = {https://doi.org/10.1016/j.aop.2011.06.004},
	issn = {0003-4916},
	journal = {Annals of Physics},
	number = {10},
	pages = {2793-2826},
	title = {Schrieffer--Wolff transformation for quantum many-body systems},
	url = {https://www.sciencedirect.com/science/article/pii/S0003491611001059},
	volume = {326},
	year = {2011}}

@article{cederbaum89block,
	author = {Cederbaum, L. S. and Schirmer, J. and Meyer, H.-D.},
	date = {1989/07/07},
	doi = {10.1088/0305-4470/22/13/035},
	isbn = {0305-4470},
	journal = {Journal of Physics A: Mathematical and General},
	number = {13},
	pages = {2427},
	title = {Block diagonalisation of Hermitian matrices},
	url = {https://dx.doi.org/10.1088/0305-4470/22/13/035},
	volume = {22},
	year = {1989}}

@article{berke22transmon,
	author = {Berke, Christoph and Varvelis, Evangelos and Trebst, Simon and Altland, Alexander and DiVincenzo, David P.},
	date = {2022/05/06},
	doi = {10.1038/s41467-022-29940-y},
	id = {Berke2022},
	isbn = {2041-1723},
	journal = {Nature Communications},
	number = {1},
	pages = {2495},
	title = {Transmon platform for quantum computing challenged by chaotic fluctuations},
	url = {https://doi.org/10.1038/s41467-022-29940-y},
	volume = {13},
	year = {2022}}

@article{moskalenko22high,
	author = {Moskalenko, Ilya N. and Simakov, Ilya A. and Abramov, Nikolay N. and Grigorev, Alexander A. and Moskalev, Dmitry O. and Pishchimova, Anastasiya A. and Smirnov, Nikita S. and Zikiy, Evgeniy V. and Rodionov, Ilya A. and Besedin, Ilya S.},
	date = {2022/11/08},
	doi = {10.1038/s41534-022-00644-x},
	id = {Moskalenko2022},
	isbn = {2056-6387},
	journal = {npj Quantum Information},
	number = {1},
	pages = {130},
	title = {High fidelity two-qubit gates on fluxoniums using a tunable coupler},
	url = {https://doi.org/10.1038/s41534-022-00644-x},
	volume = {8},
	year = {2022}}

 
\appendix
\section{Circuit Hamiltonian}\label{app:circuit_H}

In the circuit design, qubits are represented by circles and labeled as $Q_j$, where $j \in \{1, 2, 3\}$. The lumped LC series symbolize the couplers, denoted by $C_r$, with $r = \{12, 23\}$, indicating the two qubits they mutually couple. The coupling strength between a qubit $Q_j$ and a coupler $C_r$ is represented as $g_{jC_r}$. Qubits interact indirectly through their shared coupler, while they can also experience direct interactions, such as capacitive coupling. The direct coupling strength between qubit $Q_i$ and another qubit $Q_j$ is denoted by $g_{ij}$.

For simplicity in our analysis, we assume that all couplers behave as harmonic oscillators, similar to transmission lines or cavity modes. In the model for Fig.~\ref{fig:zzx_circuit}, we neglect weak interactions between qubits and distant couplers as well as coupler-coupler interaction. Specifically, we set $g_{1C_{23}}/2\pi = g_{2C_{13}}/2\pi = g_{3C_{12}}/2\pi=90$ MHz  and $g_{12}/2\pi=g_{23}/2\pi=g_{13}/2\pi=9$ MHz. While the direct coupling between the outer qubits is negligible in the \texttt{IBM\_sherbrooke} layout, we assume that \( g_{13} \) is engineered to be as strong as the capacitive couplings between adjacent qubits, in order to enhance the three-body \( ZZX \) interaction. Under this assumption, the circuit Hamiltonian can be expressed as follows:
\begin{align}
\label{eq:triangle}
H_{\rm sys}= &\sum_{j} \omega_j \hat{b}_j^{\dagger}\hat{b}_j+\frac{\delta_j}{2}\hat{b}_j^{\dagger}\hat{b}_j(\hat{b}_j^{\dagger}\hat{b}_j-1) + \sum_{r}\omega_{C_r}\hat{a}_r^{\dagger}\hat{a}_r \nonumber \\
&+ \sum_{j,r}g_{jC_r}(\hat{b}_{j}-\hat{b}_{j}^{\dagger}) (\hat{a}_r-\hat{a}_r^{\dagger}) \nonumber\\
&+\sum_{i\neq j} g_{ij}(\hat{b}_{i}-\hat{b}_{i}^{\dagger}) (\hat{b}_{j}-\hat{b}_{j}^{\dagger})
\end{align}
with $\omega_{j}$ being qubit bare frequency, $\delta_j$  qubit anharmonicity, and $\omega_{C_r}$ coupler frequency.  In order to obtain the analytical Hamiltonian between qubits in the computational subset, first we restrict circuit parameters within the dispersive regime i.e. $g_{jC_r}\ll |\omega_{C_r}-\omega_j|$, and using Rotating Wave Approximation (RWA), which helps to make the Hilbert space of harmonic couplers separable from qubit subset, therefore couplers can be safely eliminated and their effect renormalizes bare qubit frequency into their dressed values $\bar{\omega}_{j}(n_j) = \bar{E}_{j}(n_j+1) - \bar{E}_{j}(n_j)$, with dressed state energies $\bar{E_j}(n_j)$ defined as:
 \begin{align}
\bar{E_j}(n_j)=n_j \omega_j+ \frac{n_j (n_j-1)}{2} \delta_j  -\sum_r \frac{g^2_{jC_r} n_j}{\Delta_{C_r j}(n_j-1) },  \nonumber
\end{align}
with $r$ summing over  those resonators that  interact with  qubit $Q_j$ via $g_{jC_r}$, and $\Delta_{C_{r} j}(n_j)\equiv\omega_{C_r}-\omega_j-n_j\delta_j$. 

Two qubits $Q_i$ and $Q_j$ that share the same coupler $C_r$ and couple to it by the strengths $g_{iC_r}$ and $g_{jC_r}$, effectively interact with one another by the following effective strength:
\begin{align} \label{eq. Jeff}
\mathcal{J}_{ij}=g_{ij}-\frac{g_{iC_r}g_{jC_r}}{2}\left(\frac{1}{\Delta_{C_ri}(m_i)}+\frac{1}{\Delta_{C_rj}(n_{j})}\right),
\end{align}
with $g_{ij}$ being direct coupling strength between the two qubits and the second term being the perturbative indirect coupling strength via the shared coupler.

By block-diagonalizing the total Hamiltonian, the off-diagonal Pauli coefficients are evaluated perturbatively to first order in the drive amplitudes $\Omega_j$. To further streamline the expressions, we introduce an overline notation to distinguish detunings and interactions involving non-computational states: interactions confined to the computational subspace are denoted as $J_{i j}$, those involving higher excited states as $J_{\overline{i j}}$, and mixed-state interactions as $J_{i\overline{j}}$. This notation extends naturally to higher excitation levels; for instance, interactions involving the third excited state are marked with a double overline, such as $J_{\overline{\overline{i}} j}$. In particular, qubit-qubit detunings are denoted with overlines when they correspond to transitions outside the computational subspace, e.g., $\Delta_{\overline{i}j}$ or $\Delta_{\overline{\overline{i}}j}$, more details can be found in Ref.~\cite{xu25surface-code}.

In this analysis, we assume $\Omega_2 \ll \Omega_1 (\Omega_3)$, reflecting the fact that the target drive primarily serves to cancel classical crosstalk~\cite{sheldon16procedure,ku20suppression}. Under these conventions, the perturbative expressions to the first order in $\Omega$ for all relevant Pauli terms are given by:
\begin{widetext}
\begin{align}
\alpha_{ZIX} =~& \frac{1}{2}\left[
\Omega_1\cos\phi_1\left(\frac{J_{\overline{1}2}}{\Delta_{\overline{1}2}}-\frac{J_{12}}{\Delta_{12}}+\frac{J_{1\overline{3}}J_{\overline{3}2}}{\Delta_{13}\Delta_{\overline{3}2}}-\frac{J_{\overline{13}}J_{\overline{3}2}}{\Delta_{1\overline{2}}\Delta_{\overline{3}2}}\right) + \Omega_3\cos\phi_3\left(\frac{J_{1\overline{3}}J_{12}}{\Delta_{\overline{3}2}\Delta_{12}}-\frac{J_{\overline{1}\overline{3}}J_{\overline{1}2}}{\Delta_{\overline{3}2}\Delta_{\overline{1}2}}\right)
\right] \nonumber \\
&+ \frac{\Omega_2}{4}\cos\phi_2\left[
\left(\frac{J_{\overline{1}2}}{\Delta_{\overline{1}2}}\right)^2 + \left(\frac{J_{1\overline{2}}}{\Delta_{1\overline{2}}}\right)^2 - 2\frac{J_{12}J_{1\overline{2}}}{\Delta_{12}\Delta_{1\overline{2}}}
\right]\\
\alpha_{ZIY} =~&\frac{1}{2}\left[
\Omega_1\sin\phi_1\left(\frac{J_{\overline{1}2}}{\Delta_{\overline{1}2}}-\frac{J_{12}}{\Delta_{12}}+\frac{J_{1\overline{3}}J_{\overline{3}2}}{\Delta_{13}\Delta_{\overline{3}2}}-\frac{J_{\overline{13}}J_{\overline{3}2}}{\Delta_{1\overline{2}}\Delta_{\overline{3}2}}\right) + \Omega_3\sin\phi_3\left(\frac{J_{1\overline{3}}J_{12}}{\Delta_{\overline{3}2}\Delta_{12}}-\frac{J_{\overline{1}\overline{3}}J_{\overline{1}2}}{\Delta_{\overline{3}2}\Delta_{\overline{1}2}}\right)
\right] \nonumber \\
&+ \frac{\Omega_2}{4}\sin\phi_2\left[
\left(\frac{J_{\overline{1}2}}{\Delta_{\overline{1}2}}\right)^2 + \left(\frac{J_{1\overline{2}}}{\Delta_{1\overline{2}}}\right)^2 - 2\frac{J_{12}J_{1\overline{2}}}{\Delta_{12}\Delta_{1\overline{2}}}
\right]\\
\alpha_{IZX} =~& \frac{1}{2}\left[
\Omega_3\cos\phi_3\left(\frac{J_{\overline{3}2}}{\Delta_{\overline{3}2}}-\frac{J_{32}}{\Delta_{32}}+\frac{J_{1\overline{3}}J_{\overline{1}2}}{\Delta_{1\overline{2}}\Delta_{\overline{3}2}}-\frac{J_{\overline{13}}J_{\overline{1}2}}{\Delta_{\overline{1}2}\Delta_{\overline{3}2}}\right) + \Omega_1\cos\phi_1\left(\frac{J_{\overline{1}3}J_{32}}{\Delta_{\overline{1}2}\Delta_{32}}-\frac{J_{\overline{1}\overline{3}}J_{\overline{3}2}}{\Delta_{\overline{1}2}\Delta_{\overline{3}2}}\right)
\right] \nonumber\\
&+ \frac{\Omega_2}{4}\cos\phi_2\left[
\left(\frac{J_{\overline{3}2}}{\Delta_{\overline{3}2}}\right)^2 + \left(\frac{J_{3\overline{2}}}{\Delta_{3\overline{2}}}\right)^2 - 2\frac{J_{32}J_{3\overline{2}}}{\Delta_{32}\Delta_{3\overline{2}}}
\right]\\
\alpha_{IZY} =~& \frac{1}{2}\left[
\Omega_3\sin\phi_3\left(\frac{J_{\overline{3}2}}{\Delta_{\overline{3}2}}-\frac{J_{32}}{\Delta_{32}}+\frac{J_{1\overline{3}}J_{\overline{1}2}}{\Delta_{1\overline{2}}\Delta_{\overline{3}2}}-\frac{J_{\overline{13}}J_{\overline{1}2}}{\Delta_{\overline{1}2}\Delta_{\overline{3}2}}\right) + \Omega_1\sin\phi_1\left(\frac{J_{\overline{1}3}J_{32}}{\Delta_{\overline{1}2}\Delta_{32}}-\frac{J_{\overline{1}\overline{3}}J_{\overline{3}2}}{\Delta_{\overline{1}2}\Delta_{\overline{3}2}}\right)
\right] \nonumber\\
&+ \frac{\Omega_2}{4}\sin\phi_2\left[
\left(\frac{J_{\overline{3}2}}{\Delta_{\overline{3}2}}\right)^2 + \left(\frac{J_{3\overline{2}}}{\Delta_{3\overline{2}}}\right)^2 - 2\frac{J_{32}J_{3\overline{2}}}{\Delta_{32}\Delta_{3\overline{2}}}
\right]\\
\alpha_{IIX} =~& \frac{1}{2}\left[
\Omega_2\cos\phi_2 - \Omega_1\cos\phi_1\left(\frac{J_{\overline{1}2}}{\Delta_{\overline{1}2}}-\frac{J_{\overline{13}}J_{\overline{3}2}}{\Delta_{\overline{1}2}\Delta_{\overline{3}2}}\right) - \Omega_3\cos\phi_3\left(\frac{J_{\overline{3}2}}{\Delta_{\overline{3}2}}-\frac{J_{\overline{13}}J_{\overline{1}2}}{\Delta_{\overline{1}2}\Delta_{\overline{3}2}}\right)
\right] \\
\alpha_{IIY} =~&\frac{1}{2}\left[
\Omega_2\sin\phi_2 - \Omega_1\sin\phi_1\left(\frac{J_{\overline{1}2}}{\Delta_{\overline{1}2}}-\frac{J_{\overline{13}}J_{\overline{3}2}}{\Delta_{\overline{1}2}\Delta_{\overline{3}2}}\right) - \Omega_3\sin\phi_3\left(\frac{J_{\overline{3}2}}{\Delta_{\overline{3}2}}-\frac{J_{\overline{13}}J_{\overline{1}2}}{\Delta_{\overline{1}2}\Delta_{\overline{3}2}}\right)
\right] \\
\alpha_{ZZX} =~& \frac{\Omega_1}{2}\cos\phi_1\left(
\frac{J_{32}}{\Delta_{32}}\frac{J_{13}}{\Delta_{12}} - \frac{J_{32}}{\Delta_{32}}\frac{J_{\overline{1}3}}{\Delta_{\overline{1}2}} - \frac{J_{\overline{3}2}}{\Delta_{\overline{3}2}}\frac{J_{1\overline{3}}}{\Delta_{12}} + \frac{J_{\overline{3}2}}{\Delta_{\overline{3}2}}\frac{J_{\overline{1}\overline{3}}}{\Delta_{\overline{1}2}}
\right)\nonumber\\
& + \frac{\Omega_3}{2}\cos\phi_3\left(
\frac{J_{12}}{\Delta_{12}}\frac{J_{13}}{\Delta_{32}} - \frac{J_{12}}{\Delta_{12}}\frac{J_{1\overline{3}}}{\Delta_{\overline{3}2}} - \frac{J_{\overline{1}2}}{\Delta_{\overline{1}2}}\frac{J_{\overline{1}3}}{\Delta_{3\overline{2}}} + \frac{J_{\overline{1}2}}{\Delta_{\overline{1}2}}\frac{J_{\overline{1}\overline{3}}}{\Delta_{\overline{3}2}}
\right)\\
\alpha_{ZZY} =~& \frac{\Omega_1}{2}\sin\phi_1\left(
\frac{J_{32}}{\Delta_{32}}\frac{J_{13}}{\Delta_{12}} - \frac{J_{32}}{\Delta_{32}}\frac{J_{\overline{1}3}}{\Delta_{\overline{1}2}} - \frac{J_{\overline{3}2}}{\Delta_{\overline{3}2}}\frac{J_{1\overline{3}}}{\Delta_{12}} + \frac{J_{\overline{3}2}}{\Delta_{\overline{3}2}}\frac{J_{\overline{1}\overline{3}}}{\Delta_{\overline{1}2}}
\right)\nonumber\\
& + \frac{\Omega_3}{2}\sin\phi_3\left(
\frac{J_{12}}{\Delta_{12}}\frac{J_{13}}{\Delta_{32}} - \frac{J_{12}}{\Delta_{12}}\frac{J_{1\overline{3}}}{\Delta_{\overline{3}2}} - \frac{J_{\overline{1}2}}{\Delta_{\overline{1}2}}\frac{J_{\overline{1}3}}{\Delta_{3\overline{2}}} + \frac{J_{\overline{1}2}}{\Delta_{\overline{1}2}}\frac{J_{\overline{1}\overline{3}}}{\Delta_{\overline{3}2}}
\right)
\end{align}
\end{widetext}

\begin{figure}[h]
    \centering
    \includegraphics[width=0.95\linewidth]{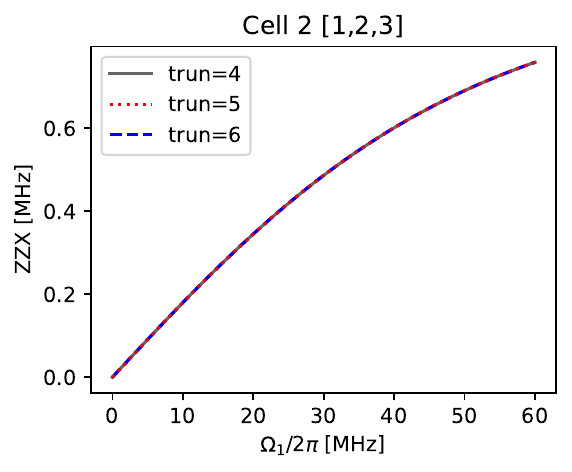}
    \vspace{-0.1in}
    \caption{$ZZX$ values for different maximum excitation numbers in unit cell 2 during GHZ state implementation.}
    \label{fig:verify_trun}
\end{figure}

Numerical simulations remain stable for maximum excitation numbers above four, as illustrated by unit cell 2 in the GHZ state implementation shown in Fig.~\ref{fig:verify_trun}.

\section{Algorithmic Framework}\label{app:algorithm}

This section describes the selection of initial tunable parameters for the optimization algorithm and the cost function definition. Throughout the protocols for GHZ state preparation, $i$Toffoli (CCNOT), and CZZ gate implementations, we fix the target interaction strength as $\alpha_{ZZX}^{\mathrm{opt}} = 0.5$~MHz. Qubit $Q_2$ is consistently the target qubit for GHZ and $i$Toffoli gates, while for the CZZ gate, $Q_1$ acts as the measurement qubit, with $Q_2$ and $Q_3$ as data qubits.

Given five control parameters, $\omega_{c_{12}}$, $\omega_{c_{23}}$, $A_1$, $A_2$, and $A_3$, the number of independent constraints does not exceed this count. For GHZ state generation, the protocol assumes $\alpha_{ZIX} \approx \alpha_{IZX} \approx -\alpha_{IIX}$ and enhances $\alpha_{ZZX}$ to $\alpha_{ZZX}^{\mathrm{opt}}$, while suppressing undesired terms. For the $i$Toffoli gate, symmetry conditions impose $\alpha_{IZX} \approx \alpha_{ZIX} \approx -\alpha_{ZZX} \approx \alpha_{IIX}$. The CCNOT gate requires $\alpha_{IZX} \approx \alpha_{ZIX} \approx \alpha_{ZZI} \approx -\alpha_{ZZX} \approx -\alpha_{IIX}$. For the CZZ gate, the key relation is $\alpha_{IZX} = -\alpha_{ZZX}$, with other terms suppressed.

Representative initial parameters satisfying these conditions for unit cells 2 ([1, 2, 3]) and 3 ([2, 3, 4]) are summarized in Table~\ref{tab:initial_params}.

\begin{table*}[ht]
\centering
\caption{Initial parameters used in the optimization algorithm for GHZ state generation, $i$Toffoli and CZZ gate implementation in unit cell 2 ([1, 2, 3]), as well as for CCNOT gate realization in unit cell 3 ([2, 3, 4]).}
\vspace{0.5em}
\begin{tabular}{ccccc}
\hline\hline
& \textbf{GHZ state} & \textbf{$i$Toffoli gate} & \textbf{CCNOT gate} & \textbf{CZZ gate} \\ \hline
$\omega_{c_{12}} / 2\pi$ [GHz] &~~5.321 & ~~5.301 &~~5.301&~~5.301\\ \hline
$\omega_{c_{23}} / 2\pi$ [GHz]& ~~5.725 & ~~5.611 &~~5.003&~~5.649\\ \hline
$A_1$&~~$0.060$ & $-0.311$&~~0.174& $-0.061$\\ \hline
$A_2$ & $-0.007$ & ~~$0.015$ &~~0.010&~~$0.005$\\ \hline
$A_3$ & ~~$1.500$ & $-1.213$ &$-0.016$&$-1.500$\\ \hline\hline
\end{tabular}
\label{tab:initial_params}
\end{table*}

\begin{table}[ht]
\centering
\caption{Physical constraints of parameters.}
\vspace{0.5em}
\begin{tabular}{cc}
\hline\hline
& Constraint  \\ \hline
$\omega_{c_{12}} / 2\pi$ [GHz] & [4.9, 7.0]  \\ \hline
$\omega_{c_{23}} / 2\pi$ [GHz]& [4.9, 7.0] \\ \hline
$A_1$& $[-1.5,1.5]$ \\ \hline
$A_2$ & $[-0.1, 0.1]$   \\ \hline
$A_3$ & $[-1.5, 1.5]$ \\ \hline\hline
\end{tabular}
\label{tab:constaints}
\end{table}

\section{Validation of Pauli Coefficients}\label{app:verification}
\begin{figure*}[ht]
    \centering
    \includegraphics[width=0.99\linewidth]{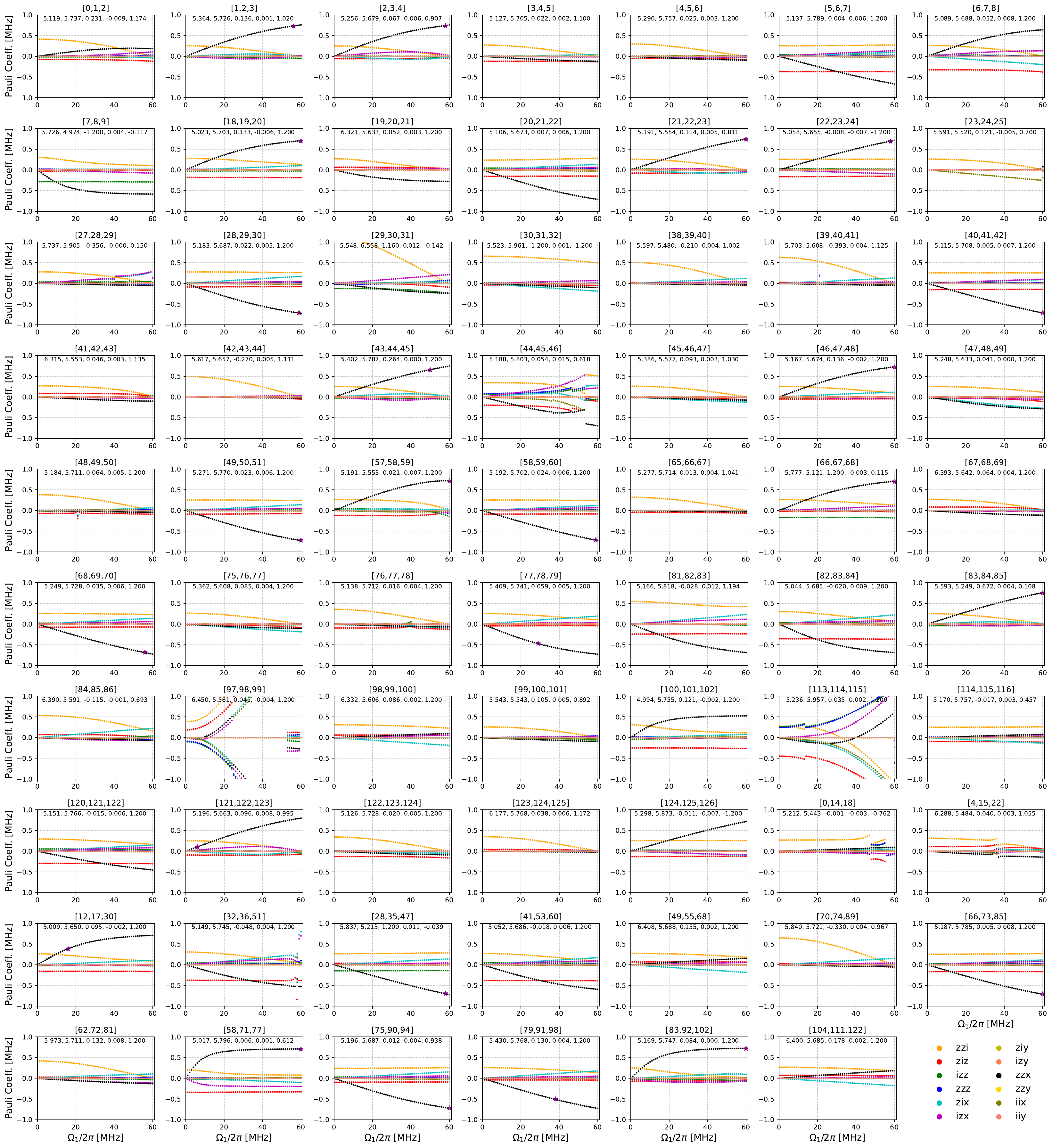}
    \caption{Verification of Pauli coefficients for GHZ state generation using optimized parameters, with $Q_1$ and $Q_3$ serving as control qubits and $Q_2$ as the target qubit. The title of each panel indicates the labels of the evaluated qubits. The inset shows the optimized parameters in the form $[  \omega_{c_{12}}, \omega_{c_{23}}, A_1, A_2, A_3]$, where $\mathrm{A}_j = \Omega_j / \Omega$, and $\Omega$ is the drive amplitude plotted along the $x$-axis. In selected panels where GHZ state fidelity exceeding 90\%, the star marks the drive amplitude at which the GHZ state reaches its maximum fidelity.}
    \label{fig:ghz_verify}
\end{figure*}

\begin{figure*}[ht]
    \centering
    \includegraphics[width=0.99\linewidth]{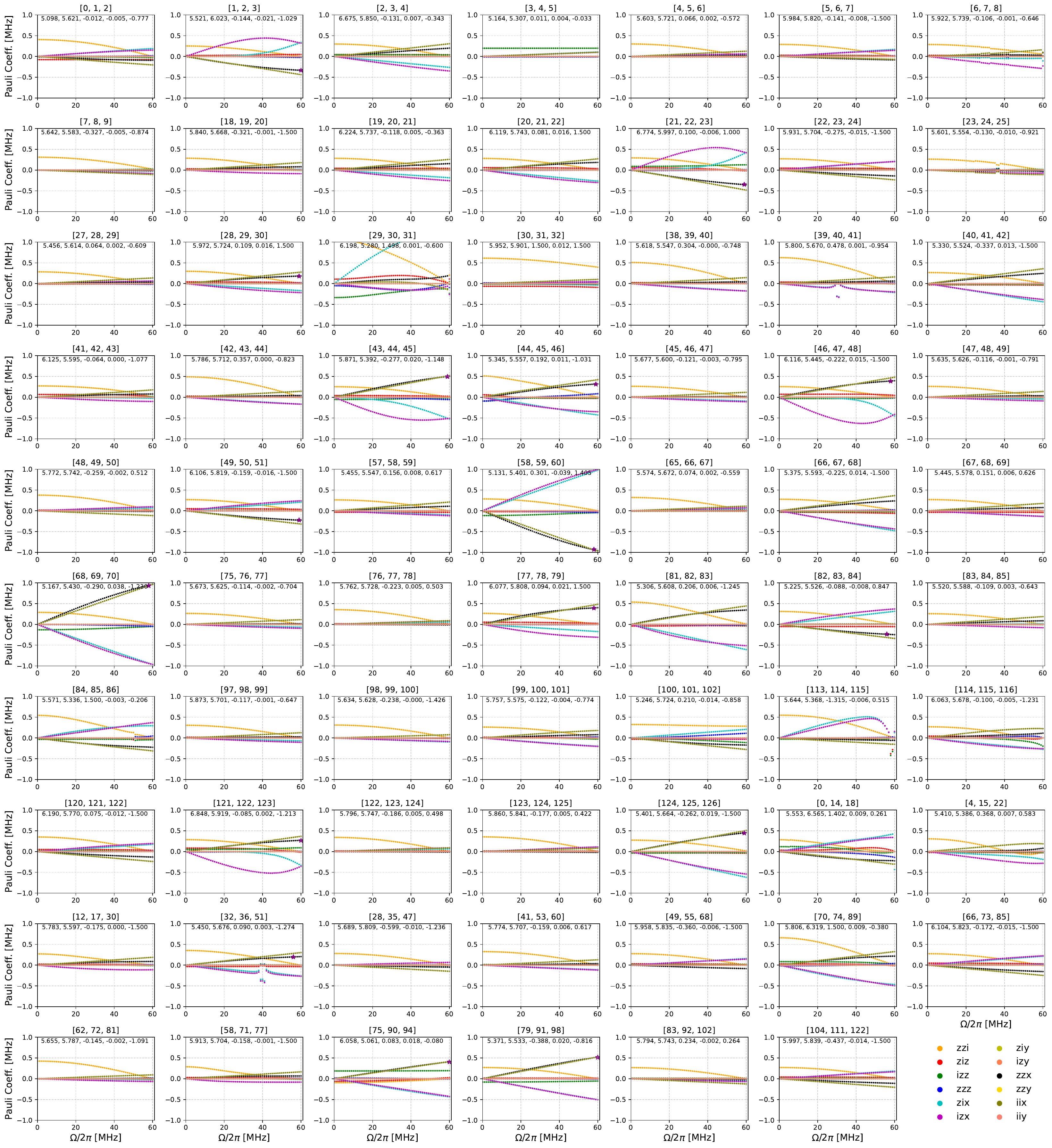}

    \caption{Verification of Pauli coefficients for generating $i$Toffoli gate using optimized parameters, with $Q_1$ and $Q_3$ serving as control qubits and $Q_2$ as the target qubit.  The title of each panel indicates the labels of the evaluated qubits. The inset shows the optimized parameters in the form $[  \omega_{c_{12}}, \omega_{c_{23}}, A_1, A_2, A_3]$, where $\mathrm{A}_j = \Omega_j / \Omega$, and $\Omega$ is the drive amplitude plotted along the $x$-axis. In selected panel where $i$Toffoli gate fidelity exceeding 90\%, the star marks the drive amplitude at which the $i$Toffoli gate reaches its maximum fidelity.}
    \label{fig:itoffoli_verify}
\end{figure*}

To confirm the effectiveness of the optimization procedure, we validate the reconstructed Pauli coefficients associated with both GHZ state generation and the $i$Toffoli gate implementation using our software CirQubit~\cite{24cirqubit}. Figures~\ref{fig:ghz_verify} and~\ref{fig:itoffoli_verify} display the extracted coefficients obtained from the nonperturbative Hamiltonian reconstruction using the optimized parameters for the GHZ state and the $i$Toffoli gate, respectively.

\begin{figure*}[ht]
    \centering
    \includegraphics[width=0.99\linewidth]{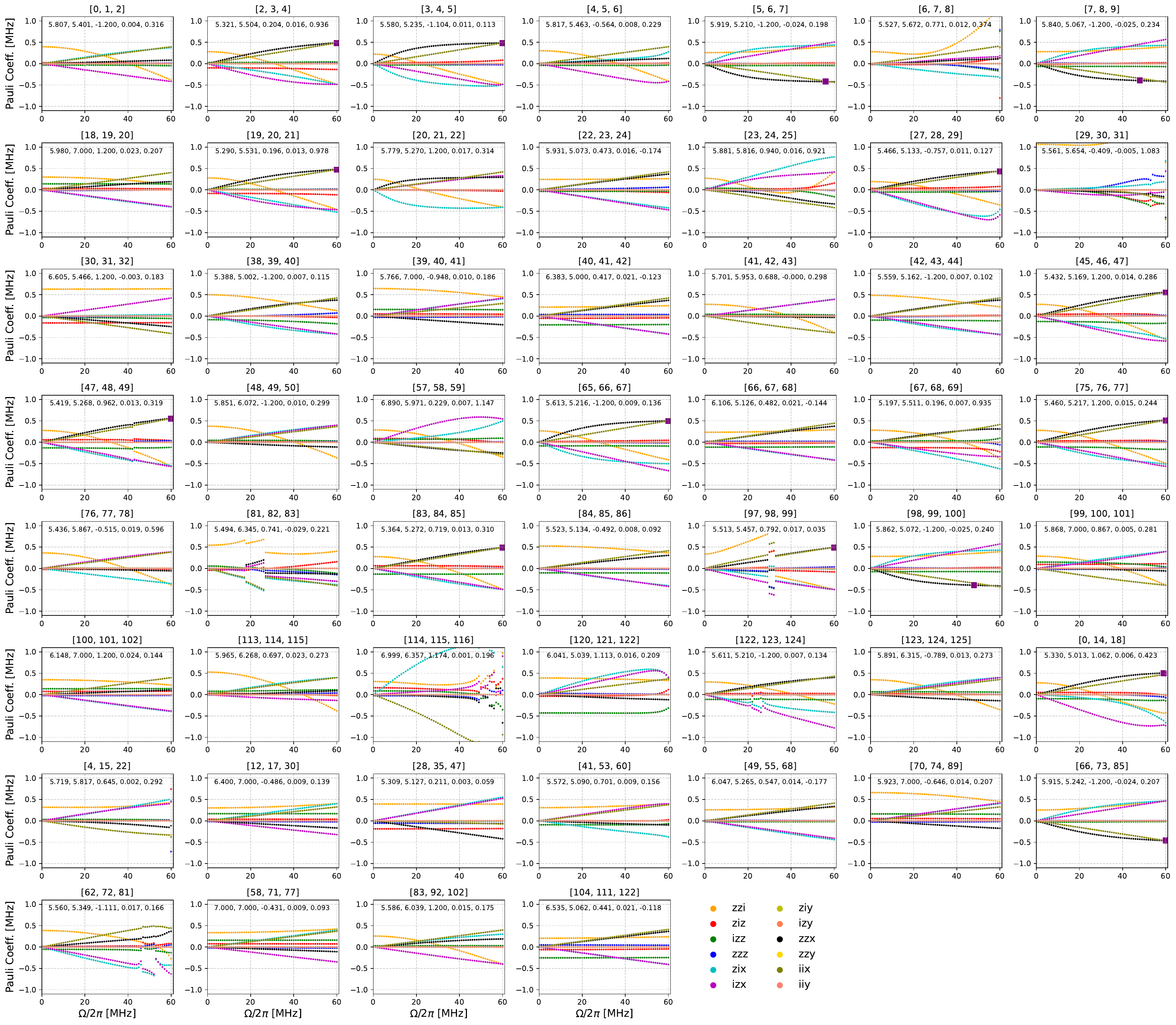}

    \caption{Verification of Pauli coefficients for generating CCNOT gate using optimized parameters, with $Q_1$ and $Q_3$ serving as control qubits and $Q_2$ as the target qubit.  The title of each panel indicates the labels of the evaluated qubits. The inset shows the optimized parameters in the form $[  \omega_{c_{12}}, \omega_{c_{23}}, A_1, A_2, A_3]$, where $\mathrm{A}_j = \Omega_j / \Omega$, and $\Omega$ is the drive amplitude plotted along the $x$-axis. In selected panel where CCNOT gate fidelity exceeding 90\%, the square marks the drive amplitude at which the CCNOT gate reaches its maximum fidelity.}
    \label{fig:ccnot_verify}
\end{figure*}

\begin{figure*}[ht]
    \centering
    \includegraphics[width=0.99\linewidth]{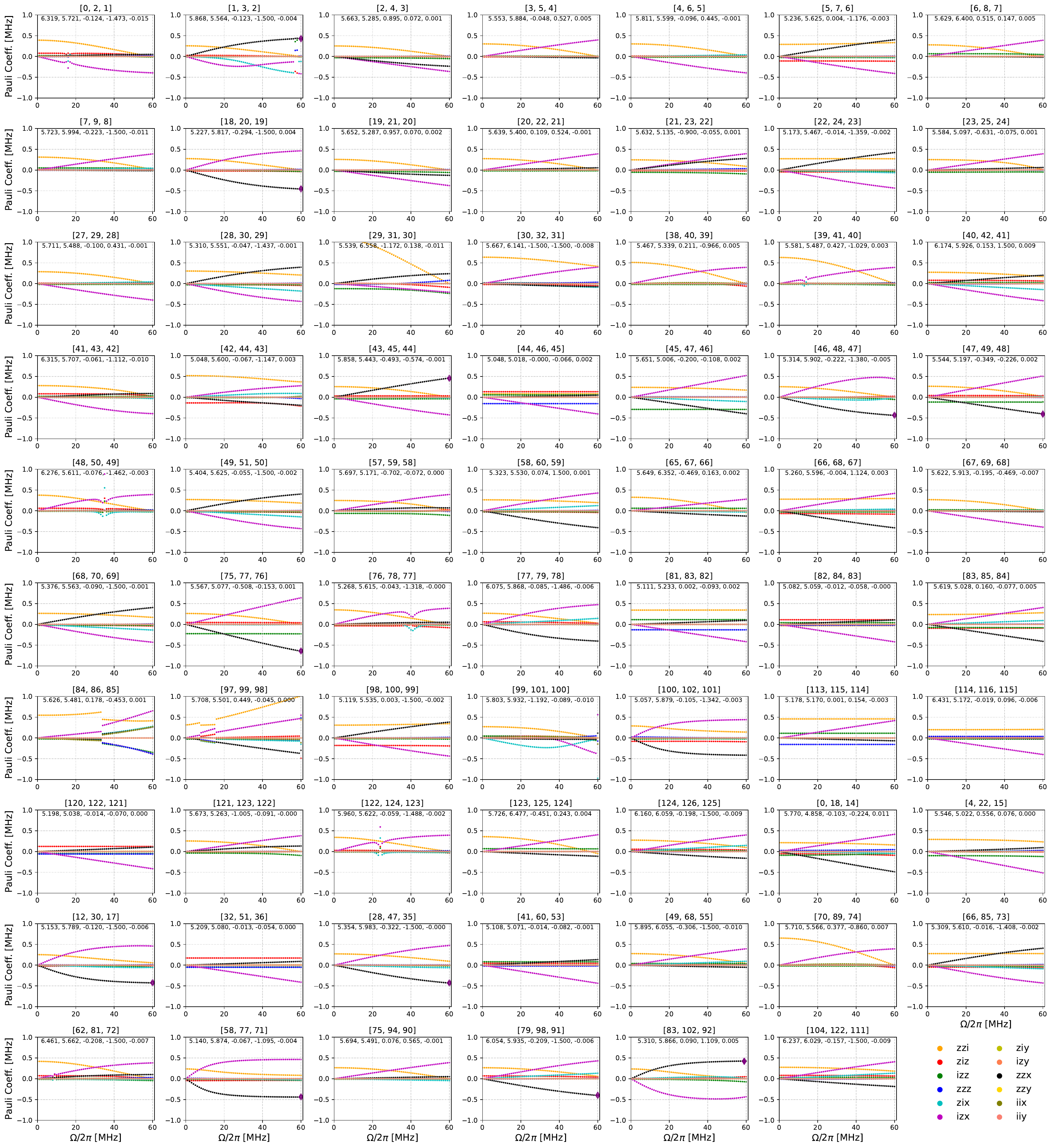}

    \caption{Verification of Pauli coefficients for generating CZZ gate using optimized parameters, with $Q_1$  serving as measure qubits, $Q_2$ and $Q_3$ as the data qubit.  The title of each panel indicates the labels of the evaluated qubits. The inset shows the optimized parameters in the form $[  \omega_{c_{12}}, \omega_{c_{23}}, A_1, A_2, A_3]$, where $\mathrm{A}_j = \Omega_j / \Omega$, and $\Omega$ is the drive amplitude plotted along the $x$-axis. In selected panel where CZZ gate fidelity exceeding 90\%, the diamond marks the drive amplitude at which the CZZ gate reaches its maximum fidelity.}
    \label{fig:czz_verify}
\end{figure*}

\end{document}